\documentclass[11pt]{article}
\usepackage{amsfonts}
\usepackage{amsmath}
\usepackage{amssymb}
\usepackage{bbm}
\usepackage{threeparttable}
\usepackage{natbib}
\usepackage{graphicx}
\usepackage{lscape}
\usepackage[justification=centering]{caption}
\usepackage{dcolumn}
\usepackage{booktabs}
\usepackage{tabularx}
\usepackage{colortbl}
\usepackage{arydshln}
\usepackage{comment}
\usepackage{todonotes}
\usepackage[hidelinks]{hyperref}
\usepackage{afterpage}
\usepackage{scrextend}
\usepackage{subcaption}
\usepackage{xcolor}
\usepackage{centernot}
\usepackage{pgf}
\usepackage{setspace}
\usepackage{url}

\presetkeys{todonotes}{size=\footnotesize}{}
\renewcommand{\baselinestretch}{1.38}
\newcommand{\lspace}[1]{\renewcommand{\baselinestretch}{#1} \small\normalsize}

\makeatother

\begin{document}
\lspace{1.0}
\vspace{-.5in}

\title{Predicted Incrementality by Experimentation (PIE) \\for Ad Measurement\thanks{\enspace We thank Neha Bhargava, JP Dub\'e, Dean Eckles, Elea Feit, Garrett Johnson, Randall Lewis, Jim Lecinski, Oded Netzer, Ilya Morozov, Kelly Paulson, Julian Runge, Ryan Shyu, Andrey Simonov, and seminar participants at Arizona State University, Carnegie Mellon University, Columbia University, Cornell University, Georgetown, HKUST, Invitational Choice Symposium, MIT, Notre Dame, Purdue University, OSU, Stanford, UNC, University of Arizona, University of Iowa, University of Maryland, University of Michigan, University of Minnesota, University of Rochester, UT Austin, Virtual Quantitative Marketing Seminar, Washington University in St.~Louis, the Utah Winter Business Economics Conference, the Conference on Digital Experimentation, the Marketing Science Conference, the NYU-Temple-CMU Conference, Amazon, Meta, and MSI.  Previously, Gordon and Zettelmeyer were part-time employees of Facebook (Academic Researchers, 3 hours/week). Currently, Gordon and Zettelmeyer hold concurrent appointments at Northwestern and Amazon. This paper describes work performed at Northwestern and is not associated with Amazon. E-mail addresses for correspondence: b-gordon@kellogg.northwestern.edu, f-zettelmeyer@kellogg.northwestern.edu, rmoakler@meta.com.}}

\author{Brett R.~Gordon \\
Kellogg School of Management\\
Northwestern University\\
\and Robert Moakler \\
Ads Research\\
Meta\\
\and Florian Zettelmeyer \\
Kellogg School of Management\\
Northwestern University and NBER\\
}
\date{April 1, 2026}

\maketitle
\thispagestyle{empty}
\lspace{1.0}

\begin{abstract}
\noindent 

Randomized controlled trials (RCTs) provide the most credible estimates of advertising incrementality but are difficult to scale. We propose Predicted Incrementality by Experimentation (PIE), which reframes ad measurement as a campaign-level prediction problem. PIE uses a sample of RCTs to learn a mapping from campaign features to causal effects, then applies it to campaigns not run as RCTs. Because the RCTs identify the causal effects, PIE can incorporate post-determined features---campaign-level aggregates such as test-group outcomes, exposure rates, and last-click conversions, computed after campaign completion. These metrics reflect the consumer behaviors that generate treatment effects, so they carry predictive information about incrementality even though they would be invalid controls in a causal model. Using 2,226 Meta ad experiments, PIE achieves an out-of-sample $R^2 = 0.88$ for incremental conversions per dollar, compared to $R^2 = 0.19$ for industry-standard 7-day last-click attribution. In a decision-making framework, PIE disagrees with RCT-based decisions in only 8--12\% of campaigns, compared to 12--20\% for last-click attribution. We conclude that PIE can help scale causal measurement from a limited number of RCTs to a large set of non-experimental campaigns.

\vspace{.2in}


\end{abstract}

\newpage
\setcounter{page}{1}

\lspace{1.2}

\section{Introduction}

Marketers are under pressure to demonstrate that digital advertising delivers benefits. The relevant causal quantity is ``incrementality''---a comparison of outcomes under an ad campaign to the counterfactual outcomes without it. Randomized controlled trials (RCTs) offer the most credible approach to measuring causal effects, but they impose substantial burdens: beyond engineering and operational complexity in auction markets, withholding ad exposures from some users creates opportunity costs for both advertisers and platforms. As a result, advertisers are reluctant to run RCTs across all campaigns \citep[e.g.,][]{johnson2023}. In practice, many rely instead on attributed metrics, such as last-click conversions, or on observational methods that exploit quasi-experimental variation in advertising exposure. While these approaches scale well, both can systematically bias causal ad effects when exposure and outcomes are jointly driven by user intent and algorithmic targeting \citep{LewisRaoReiley2011,LewisReiley2014,LewisRao2015,lift1,lift2}.\footnote{Last-click conversions attribute a conversion if the most recent ad was clicked within a predefined attribution window (e.g., 7 or 14 days). If no ad is clicked, no conversion is attributed to the campaign via click-through attribution. These counts can overestimate the causal effect of ads if ads are targeted to users who would have purchased anyway, or underestimate it if impressions that drive conversions fall outside the attribution window.}

We propose Predicted Incrementality by Experimentation (PIE), a method that uses a sample of RCTs to predict treatment effects for non-experimental campaigns. Predictors can include both pre-determined features known before the campaign runs (e.g., objective, budget, vertical) and post-determined features observed after the campaign concludes (e.g., average outcomes, exposure rate, click-through rate, last-click conversions). Post-determined features carry predictive information about incrementality because they reflect the same consumer behaviors and selection processes that generate treatment effects, even though they are not themselves causal quantities. The RCTs provide unbiased estimates of causal effects, and PIE learns how those effects covary with campaign-level observables. Because identification is handled by the experiments, endogenous campaign metrics that would be inappropriate controls in a causal model (e.g., last-click conversions) can serve as informative predictors of the treatment effects. These features remain available outside of RCTs because they can be computed from the treated group in experiments and from delivered campaign data in non-experimental settings.\footnote{Post-determined features and treatment effect estimates can share mechanical sampling variation because both are computed from the same experimental sample. We discuss this concern and why it is unlikely to materially affect our results in Section~\ref{sec:application:evaluation}.} In this sense, PIE is a prediction problem rather than a causal identification problem.

We formalize this intuition using a series of stylized models in Section~\ref{sec:pie}. We begin by motivating PIE through the fundamental problem of causal inference: for each user, potential outcomes under treatment and control cannot both be observed, so causal inference can be viewed as a missing data problem \citep{ImbensRubin2015}. Traditional approaches address this problem either by estimating average effects directly, for example with difference-in-means or inverse probability weighting, or by imputing unobserved counterfactuals under modeling assumptions.

We adopt this lens but shift the unit of analysis from the user to the campaign. For RCT campaigns, we observe average outcomes under both treatment and control; for non-experimental campaigns, we observe only the average outcome in the treated population. As a starting point, we consider the simplest version of PIE, which assumes that all treated consumers are exposed to the ad (perfect compliance in exposure). Under this assumption, the model estimates a mapping between average control and test outcomes across RCTs and applies that mapping to impute the missing average control outcome for a non-experimental campaign. This yields out-of-sample predictions of causal effects for campaigns without a randomized control group.

We next generalize the model to allow for one-sided noncompliance in ad exposure and for intermediate behaviors such as ad clicks that may causally affect the outcome. In this richer setting, post-determined features remain informative predictors because they reflect the same underlying factors that shape treatment effects, even when they are endogenous to treatment. More generally, any correlation between campaign-level observables and treatment effects can aid prediction, regardless of whether that correlation arises from causal or confounded relationships.\footnote{In a standard causal inference setting, unobserved confounders that jointly affect treatment and outcomes would violate the backdoor criterion \citep{pearl2009causality} and bias causal estimates. For PIE, these same confounders are a \textit{source} of predictive power rather than a threat to identification.} In particular, unobserved factors that simultaneously drive ad exposure, intermediate behaviors, and consumer outcomes induce cross-campaign correlations among observable metrics that a predictive model can exploit.

This logic also applies to intermediate behaviors. Even if clicking on an ad has no direct causal effect on conversions, the campaign-level click-through rate can still predict treatment effects if click decisions and conversion outcomes are shaped by some of the same unobservable factors. More generally, post-determined features can remain informative predictors even when they are endogenous, because they reflect the heterogeneity that drives variation in treatment effects across campaigns.

The strength of this predictive power depends on the underlying data generating process. We use simulations to characterize when PIE should perform better or worse relative to a benchmark that predicts the same average treatment effect for every campaign. The simulations identify six conditions under which PIE performs better, including greater heterogeneity in treatment effects, a stronger positive correlation between baseline outcomes and treatment effects, and stronger selection of consumers into exposure.

In our application, we implement PIE using a random forest trained on a dataset of 2,226 Meta ad experiments, illustrating how the approach can be operationalized with a flexible supervised learning model. We use this application both to demonstrate feasibility and to show how predictive accuracy varies across campaign types and segments. The overall PIE model achieves an out-of-sample $R^2$ of 0.88 for incremental conversions per dollar spent, so predictive accuracy reflects cross-campaign heterogeneity rather than variation in campaign scale. The model performs comparably across key subgroups such as conversion funnel levels (lower, mid, upper), prospecting versus retargeting campaigns, and industry verticals. PIE also achieves higher out-of-sample $R^2$ than last-click conversions across these subgroups. Its performance improves substantially as the number of available training RCTs increases. Training with 50 RCTs yields $R^2 = 0.37$, increasing to $R^2 = 0.72$ with 400 RCTs and $R^2 = 0.81$ with 1,600 RCTs, approaching the full-sample performance of $R^2 = 0.88$.

We investigate how accuracy degrades when PIE extrapolates. The model's predictions generalize well to new campaigns by the same advertiser ($R^2 = 0.87$) but less well to advertisers absent from the training data ($R^2 = 0.72$). We examine model performance when a particular segment of advertisers is absent from the training data. Across six segmentation variables---including advertiser size, vertical, campaign year, and audience type---PIE predicts less accurately when extrapolating than when training includes the focal segment, but the degree of degradation varies substantially. For example, extrapolating from retargeting to prospecting campaigns reduces $R^2$ by only 0.9 percentage points, whereas extrapolating to travel advertisers from non-travel advertisers reduces $R^2$ by 25 percentage points and extrapolating across campaign years reduces it by 21.3 percentage points.

Finally, the purpose of measurement is ultimately to inform decisions. Improved prediction accuracy has limited practical value unless it changes decisions. To create a decision-making framework, we adopt a simple perspective drawn from industry practice in which marketers consider a campaign successful if its performance exceeds some threshold; otherwise, the campaign is viewed as unsuccessful, and the marketer would have preferred to allocate the campaign budget to another business activity with greater expected returns. We use this framing to quantify the disagreement probability between decisions made using RCTs and those using predictive models. Specifically, we quantify how often decisions based on PIE or last-click would disagree with decisions based on RCTs. We find that PIE exhibits low disagreement rates and a better balance of Type I and Type II errors than last-click attribution, indicating that even when prediction errors remain, managers using PIE would more often reach the same go/no-go decisions as with RCTs.

Methodologically, we make two related contributions. First, we reframe non-RCT measurement as a campaign-level prediction problem (as opposed to a traditional causal inference problem), treating each RCT as a labeled observation and learning a mapping from campaign features to causal effects. Second, we introduce post-determined features and formalize the conditions under which they carry information about incrementality, showing that variables often dismissed as biased proxies for causal effects can be valid and informative predictors. To the best of our knowledge, we are the first to propose the PIE approach. Although our focus is on using PIE for ad measurement, the methodology could be generalized to other types of marketing interventions.\footnote{Our approach differs from the industry ``calibration factor'' practice that scales attributed metrics by a single ratio from one RCT and reuses it thereafter \citep[p.~15]{Google_ModernMeasurement_2024}. Our formalization explains when and why this mapping works, generalizing beyond a single scaling factor by learning campaign-specific corrections.} 

Empirically and managerially, we offer a proof-of-concept using a large dataset of campaigns from a major advertising platform. PIE achieves higher predictive accuracy than common attribution benchmarks. We also propose a decision-theoretic framework for evaluating model-based incremental performance by focusing on the decision errors managers would make using these approaches relative to an RCT. 

Our approach has been adopted in industry practice. An implementation of our model is in production at Amazon Advertising, where it is used to calibrate machine learning attribution models---composed of post-determined features---to RCT outcomes, enabling the attribution models to assign touchpoint-level credit in Amazon's new multi-touch attribution (MTA) system \citep{lewis2025amazonadsmultitouchattribution}.\footnote{See also the announcement at Amazon unBoxed 2024, \url{https://tinyurl.com/3wetscdm}.} Meta's Incremental Attribution makes use of similar approaches to the PIE models described here---leveraging pre- and post-determined features across a large set of prior RCTs.  This model was developed independently by Meta during the same period as our research.\footnote{See slide 16 of \url{https://tinyurl.com/2kmuwj3n}.}

We caution against generalizing our specific empirical results to other prediction settings (e.g., different platforms). Instead, we view these results as a proof-of-concept demonstrating that the approach can work using actual RCT data from a major advertising platform. In practice, an ad platform planning to implement PIE would need to conduct out-of-sample validation tests to determine whether PIE performs sufficiently well for its intended purpose. As with any prediction problem, our approach relies on having a representative training sample and assumes the mapping between features and outcomes is stable. Platforms can design their own RCT data generation process to help ensure that these conditions are met.

The remainder of the paper proceeds as follows. Section~\ref{sec:lit} discusses our work's relationship to several literatures. Section~\ref{sec:pie} develops the formal model and sources of predictive power. Section~\ref{sec:application} introduces the application of PIE to ad experiments at Meta, and Section~\ref{sec:results} presents results and robustness checks, including subgroup analyses and tests of generalizability. Section~\ref{sec:decision_making} evaluates alignment with managerial decisions. Section~\ref{sec:conclusion} offers concluding remarks and guidance for deploying and maintaining PIE at platform scale.

\section{Relationship to Existing Literature} \label{sec:lit}

Our work relates to five distinct areas of research: advertising effectiveness, methods that combine experimental and observational data, causal machine learning and uplift modeling, transportability and external validity, and meta-analysis. We discuss each in turn, emphasizing how PIE differs where appropriate.

\paragraph{Advertising Effectiveness.} Our work builds on the literature measuring advertising effectiveness; see \citet{gordonIneffec2021} and \citet{johnson2023} for recent reviews. This literature finds that ad effects are typically small and that detecting them requires careful experimental design and large samples \citep{LewisRaoReiley2011,LewisReiley2014,LewisRao2015}. In the absence of RCTs, researchers have exploited quasi-experimental variation in ad auctions \citep{rafieian_yoga_2022,gui_nair_niu_2022}, and platforms have improved experimental practice through eligibility-based designs such as ghost ads \citep{ghostads}. These advances are important because even high-quality observational data, combined with sophisticated machine learning methods, can fail to recover incrementality when exposure and outcomes are jointly determined by user intent and targeting \citep[e.g.,][]{lift1,lift2}. At the same time, conducting RCTs for all campaigns is resource intensive: beyond engineering complexity, withholding ad exposure (or impression opportunities) imposes opportunity costs that make continuous testing difficult to scale for many advertisers.

Against this backdrop, PIE reframes measurement as a campaign-level prediction problem trained on experimental data. Rather than modeling user-level counterfactuals, we treat each RCT as a labeled observation and learn a mapping from campaign features---both pre-determined configuration settings and post-determined metrics---to causal ad effects. We then apply this learned mapping to non-experimental campaigns without requiring control group data in production. Variables that would normally be invalid as post-treatment controls in user-level causal models (e.g., last-click conversions) can be valid predictors in a campaign-level prediction task \citep[see][on ``bad controls'' and post-treatment variables]{angrist_pischke_2009,hernan_robins_2020}.

\paragraph{Combining Experimental and Observational Data.} PIE is related to work that blends experimental and observational data, though it differs in its target and granularity. One stream of research calibrates marketing mix models (MMMs) to lift tests, using experiments as priors or targets in Bayesian MMMs \citep{runge_mmm,robyn_meta,google_bayesian_mmm}. Another uses surrogate outcomes to link short-run metrics to long-run targets via experiments \citep{AtheyEtAl2019,yang_targeting_2023}. PIE shares this experiment-anchored approach but applies it to campaign-level causal prediction rather than channel-level calibration or long-run forecasting. Amazon's multi-touch attribution system provides another related example, with PIE extrapolating RCT effects across campaigns before the system allocates attribution credit to individual touchpoints \citep{lewis2025amazonadsmultitouchattribution}.

\paragraph{Causal Machine Learning and Uplift Modeling.} PIE relates to causal machine learning and marketing uplift modeling. Methods in this area use experimental data to estimate heterogeneous treatment effects at the individual level and to rank or target users by predicted incremental response \citep{athey_imbens_2016,AtheyWager2018,Rzepakowski_Jaroszewicz_2012,ascarza2018}. PIE shares the logic of using experiments to supply labels and machine learning to generalize, but differs in unit of analysis (campaigns rather than users) and objective (aggregate effects for measurement rather than individual effects for targeting). Related in spirit is \citet{HuangAscarzaIsraeli2024}, who pool multiple experiments to predict individual-level treatment effects for personalization. PIE differs in targeting campaign-level effects and in leveraging post-determined features for prediction, aiming for accurate out-of-sample prediction rather than causal explanation \citep{shmueli_2010}.

\paragraph{External Validity and Transportability.} PIE draws on the literature on external validity and transportability \citep{PearlBareinboim2011}. Conceptually, we treat the RCT dataset as the source domain and non-experimental campaigns as the target domain, and we assume that the mapping from campaign features to causal effects is stable across these domains. This is an invariance assumption: the conditional relationship $f(\mathbf{X}_r;\theta)$ estimated on RCT campaigns generalizes to non-RCT campaigns with similar feature distributions. We assess this assumption through out-of-sample validation. The economics literature has emphasized related challenges of external validity \citep{Allcott2015QJE,Vivalt2020JEEA}, and \citet{HotzImbensMortimer2005JoE} use observed covariates to predict treatment efficacy in new settings, which is conceptually close to PIE's use of campaign features to predict treatment effects for new campaigns. PIE also has a connection to transfer learning, in which a model trained on one task or domain is adapted to another. However, transfer learning typically involves a shift in the prediction task itself, whereas PIE performs the same task (predicting campaign incrementality) in both the training and prediction stages.

\paragraph{Meta-Analysis.} Finally, PIE can be viewed as a predictive, study-level extension of meta-analysis. Like random-effects meta-analysis, it pools results from many independent experiments while allowing for between-study heterogeneity. Traditional meta-analysis or meta-regression typically targets the grand mean effect and its dispersion or estimates a small set of moderator coefficients \citep{StanleyDoucouliagos2012}. PIE departs from this tradition by aiming to generate accurate out-of-sample \textit{point predictions} for individual new campaigns, leveraging the full heterogeneity in the RCT sample to map campaign-specific features
to campaign-specific causal effects.

\section{Model of Predicted Incrementality by Experimentation}\label{sec:pie}

We introduce our methodology from the perspective of an advertising platform with access to a large and representative sample of RCTs. The platform seeks to provide estimates of incremental ad effects for campaigns that were not implemented as RCTs. PIE uses the RCT dataset to train a predictive model at the campaign level that enables calculation of incremental effects for these non-experimental campaigns.

Section~\ref{sec:model_prelim} begins with model preliminaries that specify user-level behavior in the context of a collection of ad campaigns and relate these quantities to their aggregated campaign-level counterparts. In Section~\ref{sec:model_perfect_compliance}, we introduce the PIE approach under the assumption of perfect compliance in treatment assignment, meaning that all users in the test group are exposed to ads while no users are exposed in the control group. This assumption allows us to communicate the intuition and mechanics of the PIE model. Section~\ref{sec:A realistic DGP} generalizes the model to allow for one-sided imperfect compliance with selection into exposure, which provides a more realistic description of digital ad experiments. Section~\ref{sec:pie_performance} presents simulation results that characterize the conditions under which the PIE model performs better or worse. Section~\ref{sec:pie_general} generalizes the PIE model from the potential outcomes framing in Sections~\ref{sec:model_perfect_compliance} and \ref{sec:A realistic DGP} to an arbitrary supervised learning model $f(\mathbf{X}_r; \theta)$ that maps a vector of pre- and post-determined campaign features to the causal target.

\subsection{Model Preliminaries}\label{sec:model_prelim}

We observe a collection of $R$ independent campaigns indexed by $r = 1, \ldots, R$ that were implemented as RCTs. Let $\mathcal{R} = \{1, \ldots, R\}$ denote this set of RCT campaigns. We make the standard assumptions of no interference between units (SUTVA) and no spillover effects across campaigns.

Within each RCT $r$, users are indexed by $i = 1, \ldots, N_r$ and are randomly assigned to either a test group (denoted by $Z_i^{r}=1$) or a control group ($Z_i^{r}=0$) with sizes $N_{tr}$ and $N_{cr}$, respectively.

Assignment to the test group enables the possibility of ad exposure, while control group users are prevented from seeing the ad. We denote actual exposure by $D_i^{r} = 1$ (exposed) and $D_i^{r} = 0$ (not exposed). In general, $D_i^r$ and $Z_i^r$ may differ due to one-sided noncompliance, which we address in Section~\ref{sec:A realistic DGP}.

Each RCT aims to measure the impact of the campaign on the observed outcome metric $Y^{r}_i \in \mathbb{R}$, such as purchasing a product from the advertiser. We adopt the potential outcomes framework such that $Y_i^{r}(d)$ denotes the potential outcome for user $i$ in RCT $r$ under exposure status $d \in \{0,1\}$.

A user's baseline outcome (without exposure) is $\alpha_i^r = Y_i^r(0)$, and the user-specific treatment effect is $\tau_i^r = Y_i^r(1) - Y_i^r(0)$. The observed outcome is thus
\begin{align} \label{eq:simple_outcomes}
Y_i^r = Y_i^r(D_i^r) = \alpha_i^r + \tau_i^r D_i^r.
\end{align}

For each RCT $r$, let $\tau_r$ and $\alpha_r$ denote the campaign-level average treatment effect (ATE) and average baseline outcome, respectively. Across campaigns these parameters are heterogeneous, drawn respectively from distributions $G_{\tau}$ and $G_{\alpha}$:
\begin{align}
\tau_r &= \mathbb{E}[\tau_i^r \mid \text{RCT } r] \sim G_{\tau} \quad \quad  \alpha_r = \mathbb{E}[\alpha_i^r \mid \text{RCT } r] \sim G_{\alpha}
\end{align}
We refer to $(\alpha_r, \tau_r)$ as \textit{structural parameters} because they are population-level counterparts to parameters that describe user-level behavior. We assume that the advertiser, and by extension the ad platform, seeks to estimate each $\tau_r$. 

\subsection{PIE with Perfect Compliance} \label{sec:model_perfect_compliance}

For simplicity, we begin by assuming perfect compliance in advertising exposure such that all test users are exposed and no control users are exposed. Thus, exposure equals assignment ($D_i^r = Z_i^r$), so the difference-in-means estimator yields an unbiased estimate of the ATE:
\begin{align}
\hat{\tau}_r  \: & = \: \underbrace{\frac{1}{N_{tr}} \sum_{i\in\text{test}} Y_{i}^{r}}_{\overline{Y}_{tr}}  \: -  \: \underbrace{\frac{1}{N_{cr}} \sum_{i\in\text{control}} Y_{i}^{r}}_{\overline{Y}_{cr}}.
\end{align}

Figure~\ref{fig:PIE_diagram} summarizes the problem and our general solution. The problem is that we do not observe $\overline{Y}_{cr^{\prime}}$ for non-experimental campaigns $r^{\prime} \notin \mathcal{R}$---that is, counterfactual mean outcomes under nonexposure. At the campaign level, our data are complete for RCTs in that we observe average outcomes for both test and control groups, $\{ (\overline{Y}_{cr}, \overline{Y}_{tr}) \}_{r=1}^{R}$. However, for non-experimental campaigns, we have incomplete data, observing only $\overline{Y}_{tr^{\prime}}$. Our solution is to use the sample of RCTs to train a predictive model, $f(\:\cdot\:;\theta)$, at the campaign level to predict the unobserved $\overline{Y}_{cr^{\prime}}$.

The simplest PIE model estimates a regression of this form:
\begin{align} \label{eq:pie_simple_ols}
    \overline{Y}_{cr} & =  \: f\big(\:\cdot\: ;\theta\big) + \nu_r \: = \:  \theta_0 + \theta_1 \overline{Y}_{tr} + \nu_r, 
\end{align} 
where $\nu_r$ is assumed to be mean zero. In our model, $\overline{Y}_{tr}$ is an example of a \textit{post-determined feature} because it is only known after the RCT campaign is complete. We refer to $\theta$ as a predictive parameter because it does not have a causal interpretation nor does it directly describe user-level behavior as in the potential outcomes in equation~(\ref{eq:simple_outcomes}). We use a linear regression to clarify the intuition, but the underlying logic can be applied to more sophisticated machine learning models.\footnote{We use a Random Forest for our empirical application in Section~\ref{sec:results}.}  After fitting the model to obtain the estimate $\hat{\theta}$, we can generate predictions for non-experimental campaigns $r^{\prime} \notin \mathcal{R}$:
\begin{align} \label{eq:pie_simple_predict}
\overline{Y}_{cr^{\prime}}^{\textrm{pred}}  \: & =  \: \hat{\theta}_0 + \hat{\theta}_1 \overline{Y}_{tr^{\prime}} \ ,
\end{align}
where the feature $\overline{Y}_{tr^{\prime}}$ is derived from all the users in the non-experimental campaign, as there is no distinction between a test and control group. This allows us to calculate:
\begin{align}
\hat{\tau}_{r^{\prime}}  \: &=  \: \overline{Y}_{tr^{\prime}} - \overline{Y}_{cr^{\prime}}^{\text{pred}}. 
\end{align} 

\begin{figure}[t]
    \centering
    \caption{Overview of the PIE approach.}
    \usetikzlibrary{arrows.meta}

\definecolor{pptblue}{RGB}{51, 94, 150}
\definecolor{pptred}{RGB}{176, 36, 24}

\begin{tikzpicture}[
        scale=1.0,
        thick,
        node distance=2.5cm and 2cm,
        group/.style={draw, rectangle, minimum width=2cm, minimum height=1.5cm, align=center},
        dashedgroup/.style={draw, dashed, rectangle, minimum width=2cm, minimum height=1.5cm, align=center, gray},
        arrow/.style={-Stealth, thick, pptblue},
        redarrow/.style={-Stealth, thick, pptred},
        blackarrow/.style={-Stealth, thick, black}, 
        box/.style={draw, rounded corners=3.5pt, minimum height=2cm, minimum width=5cm, thick, pptblue},
        redbox/.style={draw, rounded corners=3.5pt, minimum height=2cm, minimum width=5cm, thick, pptred},
        textstyleblue/.style={pptblue, font=\sffamily\large, align=center},
        textstylered/.style={pptred, font=\sffamily\large, align=center}
      ]
      
      \node[box, label=above:{}] (container1) at (-0.5, 2.8) {};
      
      \node[group] (test) at (-1.7, 2.8) {Test \\ Group \\ Users}; 
      
      \node[group] (control) at (0.7, 2.8) {Control \\ Group \\ Users}; 
      
      \node (equation) at (-0.5, 0.9) {\Large\(\hat{\tau}_r = \overline{Y}_{tr} - \overline{Y}_{cr}\)};
      
      \draw[arrow] (test.south) -- (-1, 1.3);
      \draw[arrow] (control.south) -- (0.5, 1.3);
      
      \node[textstyleblue] (campaigns1) at (-4.8, 2.8) {Campaigns \\ \textbf{run} as RCTs \\ $r \in \mathcal{R}$}; 
    
      \node[redbox, label=above:{}] (container2) at (-0.5, -1.0) {}; 
      
      \node[group] (campaignusers) at (-1.7, -1.0) {Campaign \\ Users}; 
      
      \node[dashedgroup] (controlusers) at (0.7, -1.) {Control \\ Group \\ Users}; 
      
        \draw[redarrow] (campaignusers.north) -- (-1, 0.5);
      
      \node[textstylered] (campaigns2) at (-4.8, -1.0) {Campaigns \\ \textbf{not run} as RCTs \\ $r^{\prime} \notin \mathcal{R}$}; 
      
    \node[align=left] (equation2) at (5.5, 2.8) {\Large\(\Rightarrow \overline{Y}_{cr} = f(\overline{Y}_{tr}; \theta) + \nu_r\)};

      \draw[blackarrow] (equation2.south) -- (5.5, -0.6);

      \node[align=left] (newequation) at (5.5, -1) {\Large\(\overline{Y}_{cr^{\prime}}^{\text{pred}} = f(\overline{Y}_{tr^{\prime}}; \hat{\theta})\)};

    \end{tikzpicture} 
    \label{fig:PIE_diagram}
\begin{minipage}{0.86\textwidth} 
    \footnotesize
    \textit{Notes:} 
    This diagram illustrates the PIE approach in its simplest form. For campaigns run as RCTs, the causal effect of advertising, $\hat{\tau}_{r}$, is estimated as the difference between average outcomes in the test and control groups. PIE uses these paired observations across campaigns to learn a mapping, $f\big(\,\cdot\,;\,\theta\,\big)$, from test group outcomes to control group outcomes. For campaigns not run as RCTs, the control group is unobserved (dashed box). The trained function $f\big(\,\cdot\,;\,\hat{\theta}\,\big)$ predicts the counterfactual mean outcome that a control group would have produced, enabling estimation of incremental effects without running an RCT. We refer to right-hand side variables, such as $\overline{Y}_{tr}$, as \textit{post-determined features} because they are only known after the RCT campaign is complete.
    \end{minipage}
\end{figure}

This approach requires three important assumptions. First, we assume that the test group of an RCT campaign would generate the same observable features if the campaign had instead been run without an RCT, i.e., $\overline{Y}_{tr} \sim \overline{Y}_{tr^{\prime}}$. This assumption is typically satisfied because the decision to run an RCT affects only whether a control group is withheld from ad delivery. There are no changes in how the platform identifies or serves ads to targeted users. Test group users are unaware they are in an experiment and experience the same ad delivery process that would operate in a regular campaign. Second, the sample of RCT campaigns is representative of the non-experimental campaigns for which we want to predict. This assumption is more likely to hold when the platform controls experiment assignment---for instance, by running experiments in the background without advertiser involvement, thereby avoiding selection based on advertiser characteristics or campaign performance expectations. When instead advertisers self-select into running experiments, representativeness becomes an empirical question that should be assessed before deploying a PIE model (we discuss this in more detail in Section~\ref{sec:results}). Third, there is no concept shift, meaning that the relationship captured in the PIE model ($f\big(\,\cdot\,;\,\theta\,\big)$) is invariant from the sample used to estimate $\theta$ to the set of non-experimental campaigns used to generate predictions.

To understand the sources of predictive power for the model in equation~(\ref{eq:pie_simple_ols}), substitute the expression for user-level outcomes in equation~(\ref{eq:simple_outcomes}) into the group means. By the Law of Large Numbers, as the sample size of each RCT grows ($N_r \rightarrow \infty$), each group mean converges to its population expectation plus a sampling error:
\begin{align}
\overline{Y}_{tr} &\approx \mathbb{E}[Y_{i}^{r} \mid D_i^r = 1] + \varepsilon_{tr} = \alpha_r + \tau_r + \varepsilon_{tr}, \\
\overline{Y}_{cr} &\approx \mathbb{E}[Y_{i}^{r} \mid D_i^r = 0] + \varepsilon_{cr} = \alpha_r + \varepsilon_{cr},
\end{align}
where $\varepsilon_{tr}$ and $\varepsilon_{cr}$ are independent with $\mathbb{E}[\varepsilon_{tr}] = \mathbb{E}[\varepsilon_{cr}] = 0$. Note that we rely on perfect compliance for $\overline{Y}_{tr}$ because $D_i^r = 1$ for all test users.

Substituting the above into the simple PIE model in equation~(\ref{eq:pie_simple_ols}):
\begin{align} \label{eq:why_simple_pie_works}
    \underbrace{\textcolor{red}{\alpha_r} + \varepsilon_{cr}}_{\overline{Y}_{cr}} & = \theta_0 + \theta_1 \underbrace{( \textcolor{red}{\alpha_r} + \tau_r + \varepsilon_{tr} )}_{\overline{Y}_{tr}} + \nu_r 
\end{align}
This shows that the model should have some predictive power because $\overline{Y}_{tr}$ and $\overline{Y}_{cr}$ are correlated through the common term $\alpha_r$, the average baseline probability of conversion in RCT $r$. The predictive accuracy of the model depends on how strongly the signal in $\alpha_r$ can be isolated from the other components. Heterogeneity in $\alpha_r$ plays an important role in aiding explanatory power, whereas $\tau_r$ contributes noise. In Section~\ref{sec:pie_performance}, we discuss in more detail the conditions under which PIE should perform better or worse.

\subsection{PIE with Imperfect Compliance and Intermediate Behaviors}\label{sec:A realistic DGP}

The model in the previous section assumed perfect compliance and no intermediate behaviors. We now relax both assumptions by introducing two new variables: (1) an \emph{unobservable} $U_i^r$ that generates one-sided noncompliance in exposure and (2) an \emph{observable} $C_i^r$ that represents intermediate behaviors along a customer's conversion path.

First, users are still randomly assigned to be \textit{eligible} to be exposed through $Z_i^r$, but within the test group, exposure status $D_i^r \in \{0,1\}$ is a non-random outcome. This outcome now depends on $U_i^r$, which is observed by user $i$ but unobserved by the ad platform. For example, users with higher values of $U_i^r$ may be more engaged with online browsing and thus more likely to encounter ads. These users may also have higher baseline propensities to purchase (higher $\alpha_i^r$). Such correlated unobservables are common in digital advertising (see \citealt{LewisRaoReiley2011}) and one of the reasons why many researchers advocate for RCTs to measure such effects \citep{johnson2023}.

Formally, we follow \citet{HeckmanVytlacil2005} and \citet{WaismanGordon2025} to include $U_i^r$ in the model by modifying exposure:
\begin{align} \label{eq:exp_selection}
D_i^r & = \mathbbm{1} \left \{U_i^r \geq 1 - p_r(Z_i^r) \right \},
\end{align}
where $p_r(Z_i^r)$ governs the process of selection into exposure.\footnote{Notice that in this exposure formulation, higher $U_i^r$ implies a greater likelihood of exposure. The more common form for equation~(\ref{eq:exp_selection}) is $\mathbbm{1} \left \{U_i^r \leq p_r(Z_i^r) \right \}$, which flips the interpretation of $U_i^r$.} Under standard conditions (see \citealt{WaismanGordon2025}), we can normalize $U_i^r \sim \textrm{Uniform}(0,1)$. The probability of exposure in the test group is
\begin{align}\label{eq:propensity}
\Pr \left (D_i^r=1 \middle \vert Z_i^r = 1 \right ) = p_r(Z_i^r) \equiv \phi_r \in (0, 1)\ , 
\end{align}
whereas the exposure probability in the control group is fixed at $p_r(0) = 0$.

Second, advertising platforms typically observe various \textit{intermediate behaviors} of users along their path to an eventual conversion $Y_i^r$. These include searching for a product, visiting a product's detail page, or clicking on an ad. These behaviors may depend on the unobservable $U_i^r$ and may have their own direct effect on $Y_i^r$. For example, a user who clicks on an ad may learn something about the product that increases their likelihood of purchase, independent of the direct effect of ad exposure.

For simplicity, consider a single binary intermediate behavior such that $C_i^r \in \{0,1\}$, which we interpret as user $i$'s click on an ad for campaign RCT $r$. Because users can only click on an ad if they were exposed to the ad, $C_i^r=1$ requires that $D_i^r=1$. The probability of clicking is:
\begin{align} \label{eq:prclick}
\Pr(C_i^r = 1 \mid D_i^r = 1, Z_i^r = 1, U_i^r) = \chi(U_i^r)
\end{align}
where $\chi(\cdot)$ captures the notion that clicking may depend on a user's $U_i^r$.

With these new elements in place, we continue to refer to potential outcomes as $Y(D)$ rather than $Y(Z,D,C)$. Using this notation, potential outcomes for unexposed and exposed users are:
\begin{equation} \label{eq:potential_outcomes}
\begin{aligned}
    Y_i^r(0) &= \alpha_i^r + \gamma_r U_i^r \\
    Y_i^r(1) &= Y_i^r(0) + \tau_i^r + \delta_r C_i^r
\end{aligned}
\end{equation}
where $\tau_i^r$ is the \emph{direct} effect of ad exposure, $\gamma_r$ captures unobserved heterogeneity in baseline outcomes associated with $U_i^r$, $\delta_r$ captures the direct effect of clicking on $Y_i^r$, and $C_i^r \equiv C_i^r\left(D_i^r = 1\right)$ denotes the potential click status if a user is exposed. Combining these expressions yields observed outcomes:
\begin{align*}
    Y_i^r & = Y_i^r(D_i^r) = \alpha_i^r + \gamma_r U_i^r + \tau_i^r D_i^r + \delta_r C_i^r \ .
\end{align*}

We could regress $\overline{Y}_{cr}$ on $\overline{Y}_{tr}$, as we did in equation~(\ref{eq:pie_simple_ols}), and use this model to predict $\overline{Y}_{cr^{\prime}}^{\textrm{pred}}$. However, the advertiser may be interested in other causal quantities, such as the average treatment effect on the treated (ATT) for campaign $r$:
\begin{align*}
\psi_r \;\equiv\; \mathrm{ATT}_r \;=\; \mathbb{E}\!\left[\,Y_i^r(1)-Y_i^r(0)\,\middle|\, D_i^r=1\,\right].
\end{align*}
We can estimate the ATT under standard assumptions for analyzing RCTs with one-sided noncompliance using
\begin{equation}
\label{eq:att_estimate}
\hat{\psi}_{r} \;=\; 
\frac{\overline{Y}_{tr} - \overline{Y}_{cr}}{\overline{D}_{tr}},
\qquad
\overline{D}_{tr} \;=\; \frac{1}{N_{r}^{\mathrm{t}}}\sum_{i\in \mathrm{t}} D_i^r.
\end{equation}
In this setting, exposed test-group users coincide with compliers, so equation~\eqref{eq:att_estimate} equals the local average treatment effect for compliers, or the ATT for exposed users \citep{ImbensAngrist94}.

Predictions of $\hat{\psi}_{r}$ for non-experimental campaigns could be obtained in two steps: first by fitting a PIE model and then by substituting a prediction for $\overline{Y}_{cr}$ into equation~(\ref{eq:att_estimate}). However, it is more effective to train a modified version of the original PIE model that directly predicts the causal quantity of interest\footnote{Predicting $\hat{\psi}_{r}$ directly aligns the model's objective with the causal estimand of interest. This also avoids error amplification that might inflate variances and add nonlinearity.}:
\begin{align} \label{eq:pie_modified_ols}
   \hat{\psi}_{r} & = \theta_0 + \theta_1 \overline{Y}_{tr} + \nu_r \ . 
\end{align} 

Define the direct effect of ad exposure and click rate among the exposed as:
\begin{align*}
\bar{\tau}_r^{E} \;\equiv\; \mathbb{E}\left[\tau_i^r \,\middle|\, D_i^r=1\right],
\qquad
\bar{C}_r^{E} \;\equiv\; \mathbb{E}\left[C_i^r \,\middle|\, D_i^r=1\right].
\end{align*}
The ATT (our target) can be decomposed as the sum of the direct effect of exposure and the subsequent effect of clicking among those users who click:
\begin{equation}
\label{eq:att_decomp}
\psi_r \equiv \mathrm{ATT}_r = \bar{\tau}_r^{E} + \delta_r\,\bar{C}_r^{E}.
\end{equation}

With sufficiently large samples of users within each RCT, the group mean outcomes are approximately:
\begin{align*}
\overline{Y}_{tr} &\approx \alpha_r + \gamma_r \,\mathbb{E}[U_i^r] \ \ + \underbrace{\phi_r\,\bar{\tau}_r^{E}}_{\mathbb{E}[\tau_i^r D_i^r \mid Z=1]}
\ \ + \ \ \underbrace{\phi_r\,\delta_r\,\bar{C}_r^{E}}_{\mathbb{E}[\delta_r C_i^r \mid Z=1]}
\ + \ \varepsilon_{tr},\\
\overline{Y}_{cr} &\approx \alpha_r + \gamma_r \,\mathbb{E}[U_i^r] \ \  + \ \varepsilon_{cr}.
\end{align*}
Thus the ATT estimator in equation~(\ref{eq:att_estimate}) satisfies:
\begin{align*}
\hat{\psi}_r & =  \frac{\overline{Y}_{tr}-\overline{Y}_{cr}}{\phi_r} \ 
\approx \ 
\bar{\tau}_r^{E} \ + \ \delta_r\,\bar{C}_r^{E} \ + \ \frac{\varepsilon_{tr}-\varepsilon_{cr}}{\phi_r}.
\end{align*}
Substituting into equation~\eqref{eq:pie_modified_ols} yields the population-level relationship:
\begin{equation}
\label{eq:pie_att_expected}
\underbrace{\textcolor{red}{\bar{\tau}_r^{E}} + \textcolor{red}{\delta_r\,\bar{C}_r^{E}} + \frac{\varepsilon_{tr}-\varepsilon_{cr}}{\textcolor{red}{\phi_r}}}_{\approx \hat{\psi}_{r}}
=
\theta_0 + \theta_1 \underbrace{\Big(\alpha_r + \gamma_r \mathbb{E}[U_i^r] + \textcolor{red}{\phi_r \bar{\tau}_r^{E}} + \textcolor{red}{\phi_r \delta_r \bar{C}_r^{E}} + \varepsilon_{tr} \Big)}_{\approx \overline{Y}_{tr}} + \nu_r \, .
\end{equation}
As before, the left- and right-hand sides share common structural components. Changing the target of the prediction model from $\overline{Y}_{cr}$ to $\hat{\psi}_r$ simply alters the structural parameters that determine the sources of predictive power. In equation~(\ref{eq:why_simple_pie_works}) of the full compliance model, heterogeneity in $\alpha_r$ played a critical role in aiding explanatory power, while heterogeneity in $\tau_r$ contributed noise. Here, the situation is reversed, with predictive signal driven by cross-campaign variation in $(\phi_r, \bar{\tau}_r^{E}, \delta_r, \bar{C}_r^{E})$, while dispersion in $(\alpha_r,\gamma_r)$ only enters through $\overline{Y}_{tr}$ and thus primarily contributes noise.

In practice, an advertising platform observes many campaign-level signals beyond $\overline{Y}_{tr}$, such as exposure rate, click-through rate, or last-click conversions, all of which may encode information about the structural parameters that determine treatment effects. The next section uses simulations to characterize when these features improve prediction.

\subsection{When should we expect PIE to work better or worse?}\label{sec:pie_performance}

In this section we identify the conditions under which PIE performs better or worse in predicting incremental effects $\hat{\psi}_{r}$. Above, we have already provided some intuition for the variation in structural parameters that allows PIE to be predictive of treatment effects. To explore how a more general implementation of the PIE model performs, we simulate a model of full compliance (as in Section~\ref{sec:model_perfect_compliance}), followed by a model of partial compliance that allows for unobserved heterogeneity to be correlated with selection into treatment and conversions (as in Section~\ref{sec:A realistic DGP}). We summarize when one should expect PIE to perform better or worse in predicting incremental effects. \ref{Appendix:Simulation} presents a more detailed version of this section.

\subsubsection{Overview of Simulation Setup}

Each simulation constructs the potential outcomes in equation~(\ref{eq:potential_outcomes}) for a set of campaigns $r \in \mathcal{R}$, each containing $N_r$ users. We specify the distribution across campaigns of $\alpha_r$ (baseline outcomes), $\gamma_r$ (effect of user unobservable on baseline outcomes), $\tau_r$ (treatment effect of exposure), $\text{corr}(\alpha_r, \tau_r)$, $\delta_r$ (treatment effect of clicking), the probability of exposure $\phi_r$, and users' base-level probability of clicking $\Pr(C_r)$. We use these to simulate user-level behaviors for each campaign. This yields for each campaign $r$ a user-level ad-exposure status $D_i^r \in \{0,1\}$, an ad-click $C_i^r \in \{0,1\}$, and control and test group outcomes $Y_{ci}^r\in \{0,1\}$ and $Y_{ti}^r\in \{0,1\}$. In addition, we define a last-click conversion as $LCC_i^r = Y_{ti}^r \cdot C_i^r$, indicating that user $i$ both clicked and converted in the test condition.\footnote{Since the simulation does not have a time dimension, there is no attribution window $w$ associated with a last-click conversion.}

We aggregate individual-level data at the campaign level, yielding for each campaign $r$ the number of conversions in each group ($Y_{tr}, Y_{cr}$), clicks ($C_{tr}\geq 0$, $C_{cr}=0$), last-click conversions ($LCC_{tr}\geq 0$, $LCC_{cr}=0$), and the ad-exposure rate ($\phi_{tr}\geq 0$, $\phi_{cr}=0$). 

We divide campaigns into equal-sized training and test samples. Using the training sample, we estimate a series of PIE specifications via OLS regression, progressively adding covariates until we arrive at the following full specification:
\begin{align}
\hat{\psi}_{r} & = \theta_0 + \theta_1 Y_{tr} + \theta_2 \phi_{tr} + \theta_3 C_{tr} + \theta_4 LCC_{tr} + \nu_r \ .
\end{align}
Note that the right-hand side features rely only on data from the test group. After obtaining $\hat{\theta}$ from the training sample, we generate predictions $\psi_{r}^{\text{PIE}}$ by applying the fitted model to campaigns in the test sample.

We evaluate prediction accuracy using out-of-sample $R^2$ between $\hat{\psi}_{r}$ and $\psi_{r}^{\text{PIE}}$. Higher $R^2$ indicates that the model substantially outperforms a naive baseline that predicts the mean treatment effect, $\bar{\psi}$, for every campaign; $R^2 = 0$ indicates no improvement over predicting the mean. We report $R^2$ for models with increasingly rich feature sets: total conversions only ($R^2_{Y_t}$), adding exposure fraction ($R^2_{Y_t, \phi}$), adding clicks ($R^2_{Y_t, \phi, C}$), and the full model including last-click conversions ($R^2_{Y_t, \phi, C, LCC}$). Each simulation is repeated 100 times with different random seeds, and we pool all test observations across repetitions to calculate a single $R^2$ for each scenario.


\subsubsection{Simulation Results for Full Compliance}

We begin with the simple case described in Section~\ref{sec:model_perfect_compliance}, where we assumed perfect compliance with advertising exposure and no intermediate behaviors. To obtain the equivalent of equation~(\ref{eq:why_simple_pie_works}) but directly predicting $\hat{\psi}_{r}$ instead of the control group outcome $\overline{Y}_{cr}$, we simplify equation~(\ref{eq:pie_att_expected}) assuming perfect compliance ($\phi_r=1$), no effect of an unobservable $U_i^r$ ($\gamma_r=0$), and no effect of intermediate behaviors $C_i^r$ ($\delta_r=0$). This yields
\begin{equation} \label{eq:pie_att_expected_mod}
\begin{aligned}
   \underbrace{\textcolor{red}{\tau_r} + \varepsilon_{tr} - \varepsilon_{cr}}_{\hat{\psi}_{r}=\hat{\tau}_{r}=\overline{Y}_{tr}-\overline{Y}_{cr}} =  \theta_0 \hspace{0.03in} + \hspace{0.03in} \theta_1 \underbrace{\bigl(\alpha_r + \textcolor{red}{\tau_r} + \varepsilon_{tr } \bigr)}_{\overline{Y}_{tr}}  + \nu_r
\end{aligned}
\end{equation}
This formulation suggests that heterogeneity in $\tau_r$ should play an important role in aiding explanatory power, while heterogeneity in $\alpha_r$ should contribute noise.

To simulate these effects, we specify a data-generating process with perfect compliance ($p_E=1$), heterogeneous baseline conversions across campaigns, heterogeneous treatment effects, and no effect of clicks on conversions ($\delta_r=0$). See \ref{Appendix:Simulation} for the exact simulation parameters. 

First, we vary the number of RCTs across four values (50, 100, 1,000, and 10,000), holding constant the sample size of each RCT at 10,000 users. Next, we fix the number of RCTs at 1,000 while varying the size of each RCT (1000, 5,000, 10,000 users). We find that PIE benefits from increased sample size: both more RCTs and larger RCTs improve performance by reducing estimation noise (Table~\ref{tab:sim_sample_size} on page~\pageref{tab:sim_sample_size}). For the remainder of the simulations, we use 1,000 RCTs with 10,000 users each.

Second, we vary the heterogeneity in baseline conversion probability $\alpha_r$ and in the treatment effect $\tau_r$. Consistent with the theoretical prediction above, greater variation in baseline conversion rates $\alpha_r$ harms performance because it adds noise to the predictors without providing information about $\tau_r$. We also find that greater variation in treatment effects $\tau_r$ improves $R^2$ because it creates more opportunity to outperform the naive baseline of predicting the mean (Table~\ref{tab:sim_heterogeneity} on page~\pageref{tab:sim_heterogeneity}).

Third, we analyze how the correlation between $\alpha_r$ and $\tau_r$ affects PIE's predictive accuracy. Recall from the right side of equation~(\ref{eq:pie_att_expected_mod}) that observed treatment group conversions depend on both $\alpha_r$ and $\tau_r$. We find that a positive correlation amplifies the signal in $Y_{tr}$ (Table~\ref{tab:sim_correlation} on page~\pageref{tab:sim_correlation}). A strong negative correlation means that in $Y_{tr} = \alpha_r + \tau_r + \varepsilon_{tr}$, $\alpha_r$ and $\tau_r$ nearly cancel each other out, leaving only noise in $Y_{tr}$ and eliminating PIE's predictive power.

After this exploration, we revert to $\text{corr}(\alpha_r, \tau_r)=0$ but now allow clicks to affect conversions ($\delta_r>0$). We find that under low click probability, introducing $\delta_r$ heterogeneity has minimal impact. When few users click, these features contain little information regardless of click effect variation (Panel A in Table~\ref{tab:sim_click_effects} on page~\pageref{tab:sim_click_effects}). Under high click probability, PIE performs better relative to a simple mean prediction as $\delta_r$ heterogeneity increases. The incremental contribution of click-based features grows substantially with $\delta$ heterogeneity (Panel B in Table~\ref{tab:sim_click_effects} on page~\pageref{tab:sim_click_effects}).

\subsubsection{Simulation Results for Partial Compliance}

We now simulate a data-generating process under partial compliance in exposure. Recall from equation~(\ref{eq:exp_selection}) that exposure is determined by $D_i^r = \mathbbm{1} \{U_i^r \geq 1 - p_r(Z_i^r)\}$, where $U_i^r \sim \text{Uniform}(0,1)$. The parameter $\gamma_r$ controls how strongly $U_i^r$ affects baseline conversion probability. When $\gamma_r > 0$, the same unobservable that determines exposure also affects baseline conversion rates, leading to exposed consumers who have higher baseline conversion probabilities than unexposed consumers. This is an empirical identification challenge for observational models.

First, we vary the heterogeneity of $\gamma_r$. We find that PIE performs less well with high heterogeneity in $\gamma_r$. This is analogous to increasing heterogeneity in the baseline probability $\alpha_r$: greater heterogeneity in $\gamma_r$ increases the variance of $Y_{tr}$ without proportionally increasing our ability to predict $\tau_r$ because $\gamma_r$ appears only on the right-hand side of equation~(\ref{eq:pie_att_expected}) (Panel A in Table~\ref{tab:sim_exposure} on page~\pageref{tab:sim_exposure}).

Second, we vary the intensity of selection into exposure. We find that PIE's post-determined features beyond total conversions---exposure fraction, clicks, and last-click conversions---become increasingly valuable in predicting treatment effects as exposure becomes more selected (lower $p_E$). Not only does the $R^2$ of the full PIE model increase, indicating that PIE outperforms a simple prediction of the mean treatment effect, but the full model also improves relative to a simple PIE model that uses only $Y_{tr}$ for prediction (Panel B in Table~\ref{tab:sim_exposure} on page~\pageref{tab:sim_exposure}). This improvement occurs because, as selection increases, only users with increasingly high $U_i^r$ are exposed, and post-determined features become increasingly informative about user composition, which PIE can exploit relative to predicting the mean.

Third, we find that click-based features improve predictions even when clicks have no direct causal effect on conversions ($\delta_r = 0$) (Panel C in Table~\ref{tab:sim_exposure} on page~\pageref{tab:sim_exposure}). This occurs because, under partial compliance, clicks reveal information about user composition. Users with high $U_i^r$ are both more likely to be exposed and, through the $\gamma_r$ parameter, more likely to convert. Since click probability also depends on $U_i^r$ (equation~\ref{eq:prclick}), click features serve as a proxy for the unobserved user characteristics that drive both exposure and outcomes. Intuitively, even if $\delta_r = 0$, the click event $C_i^r$ lies on a causal path from $U_i^r$ (through exposure $D_i^r$) and is therefore informative about the latent factors that jointly determine exposure and baseline conversion propensity. Conditioning on click features thus improves prediction of treatment effects by revealing information about user composition, not by capturing any direct click-to-conversion channel.

\subsubsection{Simulation Summary}\label{sec:pie_model_theory}

In summary, PIE performs best relative to predicting the mean treatment effect when: (1) treatment effects vary substantially across campaigns, (2) baseline outcomes vary modestly, (3) baseline outcomes and treatment effects are positively correlated, (4) the number and sample sizes of RCTs are large, (5) there is selection into exposure, and (6) the correlation between the selection mechanism and outcomes is similar across campaigns. PIE performs worst relative to predicting the mean treatment effect when treatment effects are homogeneous, baseline outcomes vary widely, or baseline outcomes and treatment effects are negatively correlated.

Having characterized when PIE should perform better or worse, translating these findings into a specific real-world setting is challenging absent detailed data from RCTs. Given access to such data, we can evaluate some of the constructs that affect PIE's performance. However, once such data are available, we recommend directly evaluating PIE using out-of-sample performance metrics, as we do in the following sections.

\subsection{General PIE Model} \label{sec:pie_general}

The models in Sections~\ref{sec:model_perfect_compliance} and~\ref{sec:A realistic DGP}, and the simulations in Section~\ref{sec:pie_performance}, served a specific purpose. They demonstrated, within a stylized potential outcomes framework, that campaign-level features can predict treatment effects because both the features and the causal quantities are functions of common structural parameters. The true model of user behavior is almost certainly more complex than any tractable formalization---involving additional unobservables, numerous intermediate behaviors, and interactions among them. One benefit of the PIE model is that the researcher need not correctly specify this data-generating process. Because the RCTs have already identified the causal effects, the prediction step inherits no identification burden. The researcher can include as features any campaign-level variables that the platform records for both experimental and non-experimental campaigns. The supervised learning model determines which features carry predictive information.

This motivates a general formulation. We introduce an aggregated feature vector $\mathbf X_r = \bigl(\mathbf{X}^{\mathrm{pre}}_r,\; \mathbf{X}^{\mathrm{post}}_r\bigr)$ partitioned into pre-determined variables, fixed \emph{ex\nobreakdash-ante} before the campaign runs, and post-determined variables, realized \emph{ex\nobreakdash-post} when the campaign concludes. The model becomes an arbitrary, possibly nonlinear function $f(\mathbf X_r; \theta)$. Let $\hat{\psi}_r$ denote the campaign-level causal target (e.g., ATT or a transformation). We write:
\begin{align} \label{eq:pie_att}
 \hat{\psi}_r & = f \bigl(\mathbf X_r; \theta \bigr) + \nu_r,
 \qquad \mathbb E[\nu_r \mid \mathbf X_r]=0.
\end{align}
As a special case, the one-regressor PIE model in Section~\ref{sec:model_perfect_compliance} had $\mathbf X_r = \bigl(\mathbf{X}^{\mathrm{pre}}_r,\; \mathbf{X}^{\mathrm{post}}_r\bigr) = \bigl(\emptyset,\; \overline{Y}_{tr}\bigr)$. The mean outcome in the test group, $\overline{Y}_{tr}$, is an example of a post-determined feature because it is known only after the campaign concludes and is formed by aggregating user-level variables.

More generally, post-determined features such as exposure rate, click-through rate, or last-click conversions are informative because they too depend on the behavioral and structural factors that drive treatment effects.\footnote{Because PIE is a prediction problem, the temporal ordering of features and the target variable is irrelevant for validity. Lower-funnel outcomes (e.g., purchases) are valid predictors of upper-funnel treatment effects, and vice versa, because the model exploits correlations across campaign-level aggregates rather than within-user causal pathways.} These metrics depend on actual conversion outcomes ($Y_i^r$) as in the case of last-click conversions, or on intermediate behaviors ($C_i^r$) as in the case of click-through rates, or both. We can express these features generically using:
\begin{align}
\mathbf{X}^{\mathrm{post}}_r & =
          \frac{1}{N_r}\sum_{i=1}^{N_r}
          h\bigl(Y_i^{r},\,D_i^{r},\,C_i^{r},\,\text{aux}_i^{r}\bigr)
\end{align}
where $h(\cdot)$ is a vector-valued mapping and $\text{aux}_i^{r}$ denotes any additional user-level signals (e.g., page views, ad placement indicators, invalid clicks) that the platform logs.

In addition, pre-determined features $\mathbf{X}^{\mathrm{pre}}_r$ that include campaign configuration options (e.g., objective, creative, budget, targeting options) or advertiser-level descriptors (e.g., vertical) may shift the distribution of unobservables or the shape of treatment heterogeneity, so including them in $f(\cdot)$ helps the model adapt to concept shift across advertisers. The platform need only ensure that it logs the necessary metrics from all campaigns with which to train the model.

\section{Application to Experiments at Meta}\label{sec:application}	

Having characterized the theoretical conditions under which PIE should perform well or poorly, we now turn to implementation and evaluation. In this section, we describe the dataset of Meta ad experiments to which we apply the general version of the PIE model described in Section~\ref{sec:pie_general} and explain our specific estimation and evaluation approach. Results follow in Section~\ref{sec:results}.

\subsection{Data}\label{sec:application:setting}

The data consist of ad experiments run by advertisers using Meta's ``Conversion Lift'' product, which delivers ads across Facebook, Instagram, and the Facebook Audience Network.\footnote{\url{https://www.facebook.com/business/help/221353413010930?id=547299432790676}, accessed on October 14, 2025.} Each experiment randomly assigns users in the advertiser's target audience to test (eligible for exposure) or control (ineligible) groups, after which delivery occurs via standard auctions.\footnote{For details on ad delivery and advertising RCTs at Meta, see Section 2 of \cite{lift2}.} The experiments in our data are a superset of those in \cite{lift2}.\footnote{Data requirements differed between these papers, allowing us to include experiments that were excluded from \cite{lift2}.}

We randomly selected experiments representative of large-scale tests run by U.S.~advertisers between November 1, 2019, and March 1, 2020, with at least one million test group users and at least 5,000 test group conversions.\footnote{These thresholds help balance statistical power and computational feasibility.} The resulting sample contains 839 experiments across different industry verticals with varying targeting choices, campaign objectives, conversion outcomes, and budgets. An advertiser can run multiple treatment-control pairs in a multi-cell test to evaluate different strategies. We treat each cell (treatment-control pair) as a unique experiment since populations do not overlap by construction. In our sample, 120 experiments include multiple pairs, yielding 998 unique treatment-control pairs, which we refer to as experiments going forward.

The data contain approximately 11.4 billion user-experiment observations with 52.5 billion ad impressions. The median experiment runs for 28 days, includes 7.4 million users, and delivers 21.5 million impressions. The median control group share is 10\%, and the median test group exposure rate is 76.3\%. Our sample spans many industry verticals, with the e-commerce and retail verticals comprising roughly half of all RCTs. Distributional summaries of duration, reach, splits, exposure rate, and impressions are shown in Appendix Figure~\ref{fig:dist_app}.

Most experiments measure multiple conversion outcomes (e.g., purchases, checkout initiations, app installs, page views). While advertisers choose which events to measure, these typically fall into three funnel stages: upper funnel (e.g., page views), mid-funnel (e.g., cart additions), and lower funnel (e.g., purchases). Our 998 experiments measure 2,226 conversion events across the funnel. We refer to each experiment-conversion event pair as an RCT and classify RCTs as Lower Funnel (773), Mid Funnel (713), or Upper Funnel (740).

We estimate the average treatment effect on the treated (ATT) for each RCT using the Wald/2SLS estimator with randomized assignment as the instrument \citep{ImbensAngrist94}, which under one-sided noncompliance and monotonicity identifies the average effect for exposed users. We compute the implied number of (estimated) incremental conversions as
\begin{align}
\mathrm{IC}_r & = \widehat{\text{ATT}}_{r} \cdot \overline{D}_{tr} \cdot N_{tr} \ ,
\end{align}
that is, estimated ATT times the number of exposed users in the test group.

Consistent with Section~\ref{sec:pie_general}, the features available for non-experimental campaigns must be computable without data from the control group. For post-determined features entering $\mathbf{X}^{\mathrm{post}}_r$, we use (1) the mean outcome in the test group $\overline{Y}_{tr}$; (2) the exposure rate $\overline{D}_{tr}$; and (3) the number of last-click conversions $\text{LCC}_{tr}(w)$ measured over standard attribution windows $w\in\{\text{1 hour},\text{1 day},\text{7 day},\text{28 day}\}$.\footnote{Our data do not contain raw clicks $C_i^r$ or total clicks. However, one could interpret the $\text{LCC}_{r}(w)$ as being formed from the interaction of a conversion event $Y_i^r$ and a click event $C_i^r$, constrained to occur within a window $w$ of each other.} Because the prediction target is expressed in conversions per dollar (ICPD), we normalize $\overline{Y}_{tr}$ and each $\text{LCC}_{tr}(w)$ by total campaign spend so that features and outcome are on a common scale.\footnote{The exposure rate $\overline{D}_{tr}$ is already a proportion and does not require this normalization.} We note that our data contain fewer post-determined features than a platform would typically observe.

Finally, for pre-determined features that enter $\mathbf{X}^{\mathrm{pre}}_r$, we observe advertiser- and campaign-level features for each experiment, including (1) targeting settings, (2) bidding strategy, (3) optimization settings, (4) advertiser characteristics and platform experience, and (5) actual campaign spending.

Next, we explain the specific implementation and evaluation of the model using these data.

\subsection{Model Estimation and Evaluation} \label{sec:application:evaluation}

For the outcome $\hat{\psi}_r$ of the prediction model, we use a simple transformation of the ATT, namely the number of incremental conversions per dollar (ICPD):
\begin{align} \label{eq:icpd_calculation}
\text{ICPD}_r & = \frac{\mathrm{IC}_r}{\text{Cost}_r} \ . 
\end{align}
The ICPD is a common financial metric that advertisers use to assess campaign performance and helps normalize treatment effects against different budget levels.\footnote{A related metric is the cost per incremental conversion (CPIC), the inverse of ICPD, which advertisers often use to assess profitability. Because advertisers typically have private information on the profit per conversion, CPIC helps them determine whether a campaign produced a positive return on ad spend (ROAS). From a modeling standpoint, however, CPIC is problematic because (i) it is undefined when estimated lift is zero or negative and (ii) it produces extremely skewed error distributions when lift is small. ICPD avoids both issues.}

We implement $f \bigl(\mathbf X_r; \theta \bigr)$ as a random forest and train on various combinations of the pre- and post-determined features enumerated in Section~\ref{sec:application:setting}.\footnote{We explored other models. We found that linear regression generally underperformed and that XGBoost performed similarly or slightly better than random forests but was more computationally burdensome.} We perform hyperparameter tuning for each model specification using 10-fold cross-validation.

To train and evaluate the model, we use weighted out-of-sample $R^2$:
\begin{align}
R^2 = 1 - \frac{\sum_r \omega_{r} \cdot (\text{ICPD}_{r} - \text{ICPD}_{r}^{\text{PIE}})^2}{\sum_r \omega_{r} \cdot(\text{ICPD}_{r} - \overline{\text{ICPD}})^2}
\end{align}
where $\text{ICPD}_{r}^{\text{PIE}}$ is the prediction, $\text{ICPD}_{r}$ is the observed value, $\overline{\text{ICPD}}$ is the mean observed value, and $\omega_{r}$ is the weight associated with RCT $r$, which sum to one. We use the campaign cost as the weight to reflect the notion that the ad platform prefers higher accuracy on more expensive campaigns.\footnote{We have considered weights based directly on the inverse of the ATT's sample variance, including versions based on calculations using methods from the meta-analysis literature to better account for within- and between-experiment variance. We do not find that this materially changes our results.} We use a nonparametric bootstrap with 1,000 draws to estimate the distribution of out-of-sample $R^2$.\footnote{Because $\text{ICPD}_r$ is estimated, the observed $R^2$ has a ceiling below one: even a perfect model would leave residual variance from estimation noise in the outcomes. Given our sample sizes, this noise may be small relative to the cross-campaign variation in true effects, suggesting a ceiling close to one, though we cannot verify this without decomposing the residual into model error and outcome noise. We thank Randall Lewis for raising this point.}

The preferred evaluation metric and weights are design choices for the advertising platform, and we are not wedded to the particular choices made in this empirical application. The platform could train different models using different evaluation metrics and weights, depending on their use case.

One concern with our evaluation is that the post-determined features $\overline{Y}_{tr}$ and $\overline{D}_{tr}$ also appear in the construction of the dependent variable $\text{ICPD}_r$. This shared dependence on finite-sample realizations can induce mechanical correlation between features and outcomes, potentially inflating out-of-sample $R^2$. The standard remedy is within-campaign sample splitting: using one random half of users to compute the outcome and the other half to compute features, as we implement in our simulations (\ref{Appendix:Simulation}). Our empirical application uses pre-aggregated campaign-level data, which precludes this approach. However, the large median sample size of our RCTs (7.4 million users) implies that sampling noise in both features and outcomes is small, limiting the scope for mechanical correlation to materially affect our results.

\section{Results} \label{sec:results}

This section presents baseline results and robustness evaluations from applying our PIE model to the 2,226 RCTs. We first present the model's performance across all RCTs and analyze its predictive accuracy across several campaign characteristics: purchase funnel level, prospecting versus retargeting campaigns, and industry vertical. We also examine performance as a function of the number of RCTs in the training sample.

Second, in Section~\ref{sec:results:newexisting}, we assess the performance of PIE for new campaigns of advertisers whose prior campaigns were used to train PIE. We compare this to the performance of PIE for new campaigns of advertisers without prior campaigns in PIE's training data. This examines the model's ability to transfer information across advertisers, testing how well the model can address the ``cold-start'' problem of generalizing to completely new advertisers.

Third, in Section~\ref{sec:results:segment}, we examine the model's ability to transfer information across campaign segments. For example, we assess PIE's ability to extrapolate by predicting campaign performance for large advertisers when the model is trained only on RCTs from smaller advertisers. We compare this predictive performance to the within-segment performance when the model is trained on RCTs from large advertisers.

\subsection{Baseline}\label{sec:results:baseline}

\begin{figure}[t]
    \centering
    \caption{PIE model performance.}
    \includegraphics[width=0.98\textwidth]{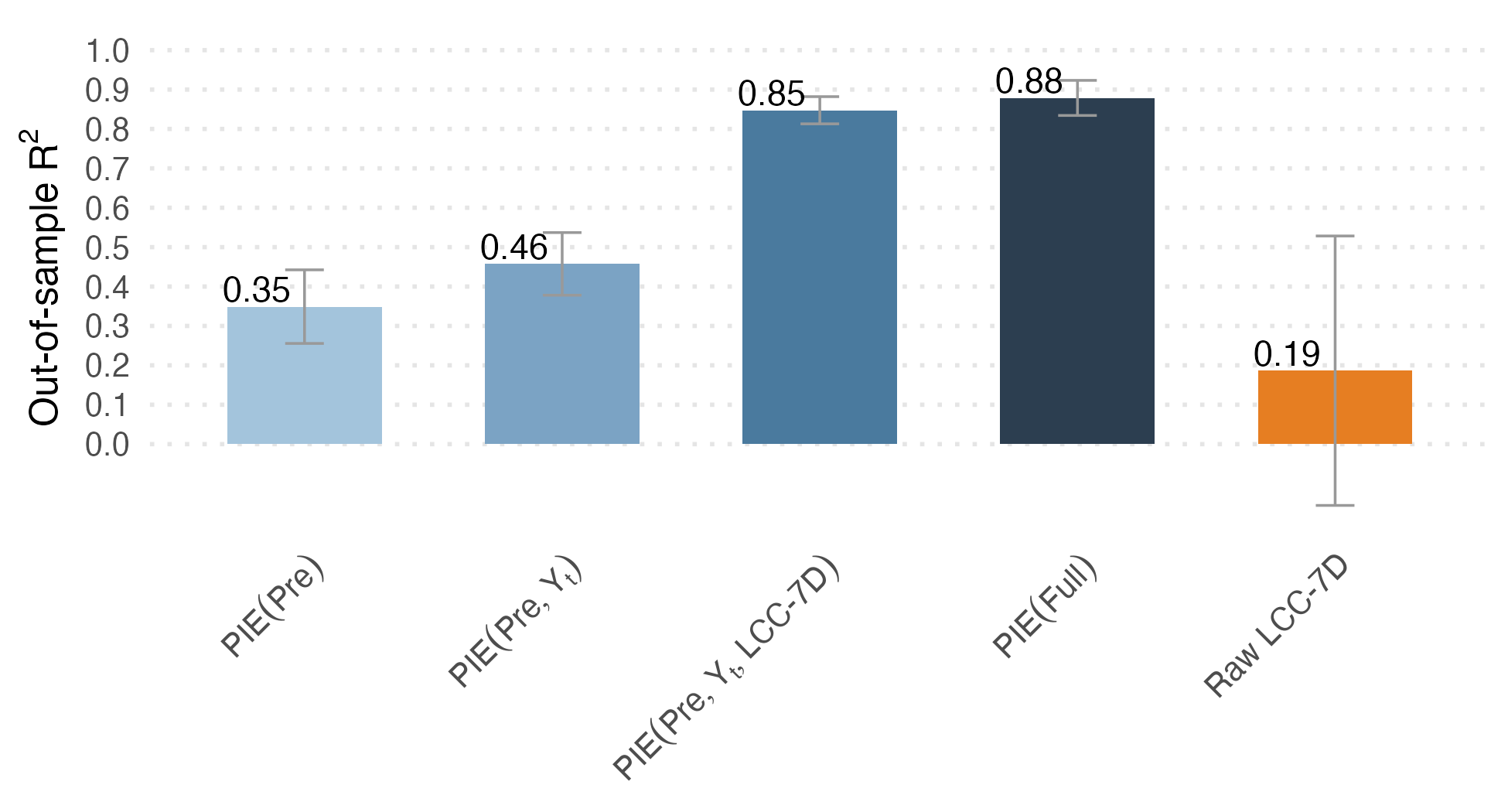} 
    \label{fig:pie_main_results}
\begin{minipage}{0.98\textwidth} 
    \footnotesize
    \textit{Notes:} This figure compares the out-of-sample $R^2$ to predict ICPD ($\hat{\psi}_{r}$). The bars represent progressive additions of features to the PIE model, from $\text{PIE}(\text{Pre})$ (using only pre-determined features) to $\text{PIE}(\text{Full})$ (using all features). $\text{Raw LCC-7D}$ serves as a benchmark that divides the raw last-click 7-day conversion count by campaign spend and treats the result as if it were ICPD, with no model applied. The $R^2$ value is shown above each bar, and error bars represent one standard deviation from the mean $R^2$, estimated via bootstrap.
    \end{minipage}
\end{figure}

Figure~\ref{fig:pie_main_results} presents the baseline performance of several PIE model specifications. $\text{PIE(Pre)}$ includes campaign characteristics (which are pre-determined variables) in $\mathbf{X}^{\mathrm{pre}}_r$ and omits any post-determined variables ($\mathbf{X}^{\mathrm{post}}_r = \emptyset$). Next, in $\text{PIE(Pre}, Y_t)$ we add one post-determined variable, such that $\mathbf{X}^{\mathrm{post}}_r = \{\overline{Y}_{tr}\}$, while we retain $\mathbf{X}^{\mathrm{pre}}_r$. $\text{PIE}(\text{Pre}, Y_t, \text{LCC-7D})$ adds $\text{LCC}_{r}(\text{7 day})$, the last-click 7-day conversion counts, and $\text{PIE(Full)}$ adds the remaining post-determined features. The error bars on the graph represent one standard deviation from the mean $R^2$.

We compare the performance of these specifications to each other and against a benchmark of $\text{Raw LCC-7D}$. This benchmark involves no model: it divides the raw last-click 7-day conversion count by campaign spend and treats the result as if it were ICPD. The resulting $R^2$ measures how well this unadjusted metric tracks true incremental effects. We focus on this attribution window because it is a common choice for advertisers.

We draw several observations from Figure~\ref{fig:pie_main_results}. First, adding the first post-determined feature, $\overline{Y}_{tr}$, to the model increases out-of-sample $R^2$ from 0.35 to 0.46 despite the model already conditioning on pre-determined features that describe a campaign's audience targeting, budget, bidding strategy, and advertiser type. Second, including LCC-7D as another post-determined feature leads to a substantial $R^2$ increase to 0.85. Adding more post-determined features in the $\text{PIE(Full)}$ specification, such as other LCC metrics, yields a modest increase to $R^2=0.88$. The additional LCC metrics likely produce relatively small gains because they are mechanically correlated with each other. More broadly, the limited marginal gains from adding other LCC windows reflect the narrow scope of the available post-determined variables in our dataset. Our data contain fewer post-determined features than a platform would typically observe. A platform designing its data collection to support PIE could incorporate a wider range of behavioral signals (e.g., view-through conversions, engagement metrics, dwell times) that might yield further improvements in predictive accuracy.

Critically, all PIE models outperform the raw LCC 7-day metric. Using the best model, an advertiser increases the $R^2$ of the prediction from 0.19 to 0.88, a substantial increase in predictive performance for treatment effects in campaigns that were not run as RCTs. The PIE predictions also exhibit lower variance compared to using last-click conversions alone. In general, out-of-sample $R^2$ values can be negative; this occurs when the predictor performs worse than simply using the mean outcome. This is particularly true for raw last-click metrics, which involve no model and simply treat the attribution count as if it directly measured incrementality.

The contrast between the strong contribution of LCC-7D within PIE and the weak standalone performance of Raw LCC-7D reflects a fundamental distinction. PIE learns across campaigns how last-click conversions relate to true incremental effects. Raw LCC-7D involves no such learning and simply assumes that attributed conversions equal incremental conversions. Across the full sample, mean LCC-7D per dollar overstates mean ICPD by a factor of 1.33. An OLS regression of $\hat{\psi}_r$ on LCC-7D per dollar yields a slope of 0.69, indicating that on average approximately 69\% of LCC-7D per dollar are incremental. Thus, using LCC-7D per dollar as a direct substitute for ICPD introduces a systematic upward bias that is reflected in the out-of-sample $R^2$. The LCC-7D metric nonetheless remains highly correlated with true incremental effects (Spearman $\rho = 0.89$), indicating that it carries substantial information about incrementality. PIE exploits this information by learning the mapping from confounded attribution metrics to causal quantities.

Comparing $\text{PIE}(\text{Pre}, Y_t)$ and $\text{PIE}(\text{Pre}, Y_t, \text{LCC-7D})$ reveals the importance of including last-click 7-day conversions as a predictive feature. The predictive contribution of a last-click metric---often viewed as a biased measure of causal ad effects---likely reflects the nature of ad exposure on social media platforms where users typically visit the platform to browse content, not to search for or purchase specific products. As a result, the endogeneity of the click event is substantially attenuated relative to search advertising. If a user stops scrolling, clicks on an ad, and then purchases within seven days, the click-to-conversion  sequence is likely informative about a genuine causal effect of the ad. In the framework of our model, this informativeness arises through two channels. First, the click event $C_i^r$ may have a direct causal effect on the outcome ($\delta_r > 0$), for instance because clicking exposes the user to additional product information. Second, absent a direct effect ($\delta_r = 0$), the click reveals information about the user's unobservable $U_i^r$, which in turn correlates with both exposure and baseline conversion propensity, as demonstrated in our simulations (Panel~C of Table~\ref{tab:sim_exposure}).

This differs from search advertising, where users arrive with pre-existing purchase intent. A user who searches for a product, clicks on the first sponsored result, and then purchases generates a last-click conversion that is heavily confounded by the intent that drove the search. In such settings, last-click conversions may be a substantially noisier signal of incremental ad effects. This distinction implies that the strong predictive power of LCC-7D documented here should not be  assumed to generalize to search advertising platforms without empirical validation.

\begin{figure}[t]
    \centering
    \caption{PIE model performance by Conversion Event Funnel}
    \includegraphics[width=0.98\textwidth]{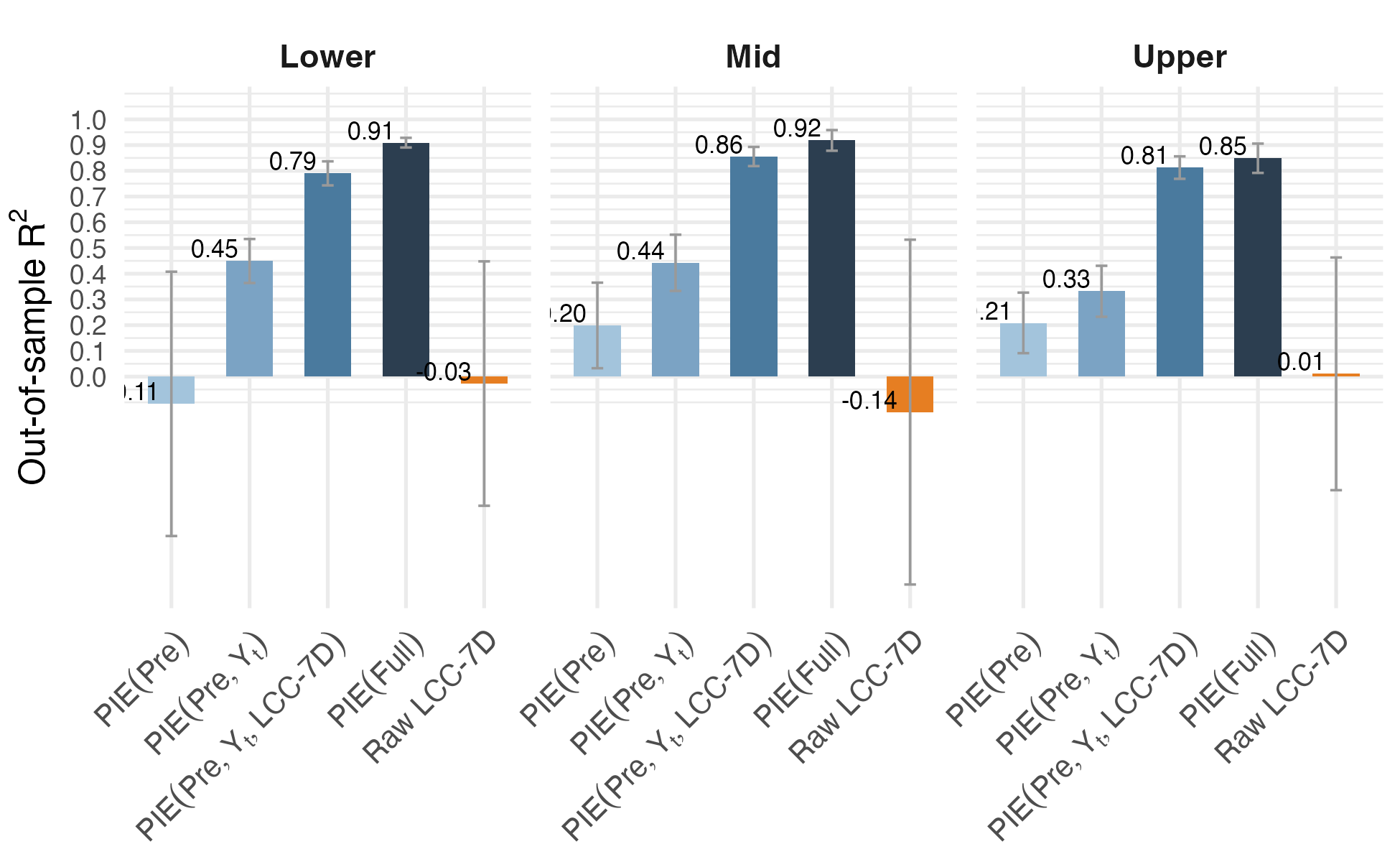} 
    \label{fig:pie_main_results_funnel_rsq}
\begin{minipage}{0.98\textwidth} 
    \footnotesize
    \textit{Notes:} This figure compares the out-of-sample $R^2$ to predict ICPD ($\hat{\psi}_{r}$), decomposed by the type of outcome according to the conversion event funnel. Lower funnel events mostly represent online purchases, whereas upper funnel conversions are most frequently page landings. The bars represent progressive additions of features to the PIE model, from $\text{PIE}(\text{Pre})$ (using only pre-determined features) to $\text{PIE}(\text{Full})$ (using all features). $\text{Raw LCC-7D}$ serves as a benchmark that divides the raw last-click 7-day conversion count by campaign spend and treats the result as if it were ICPD, with no model applied. The $R^2$ value is shown above each bar, and error bars represent one standard deviation from the mean $R^2$, estimated via bootstrap.
    \end{minipage}
\end{figure}

Figure~\ref{fig:pie_main_results_funnel_rsq} decomposes the earlier results by the level of purchase funnel of the conversion event. Across the funnel levels, we observe a pattern consistent with the baseline results: conditioning on $\overline{Y}_{tr}$ improves over using only the pre-determined variables $\mathbf{X}^{\mathrm{pre}}_r$, conditioning on $\text{LCC}_{r}(\text{7 day})$ yields even larger improvements, and adding the remaining post-determined variables produces modest gains. In this case, using raw last-click conversions (per dollar) yields negative $R^2$, indicating that a prediction based on the mean ICPD across RCTs would be more accurate than using raw last-click conversions to predict ICPD for new campaigns. The full model, in contrast, achieves $R^2$ values of 0.91, 0.92, and 0.85, for lower, mid-, and upper-funnel outcomes, respectively. 

\begin{figure}[t]
    \centering
    \caption{PIE model performance by Prospecting vs.~Retargeting}
    \includegraphics[width=0.98\textwidth]{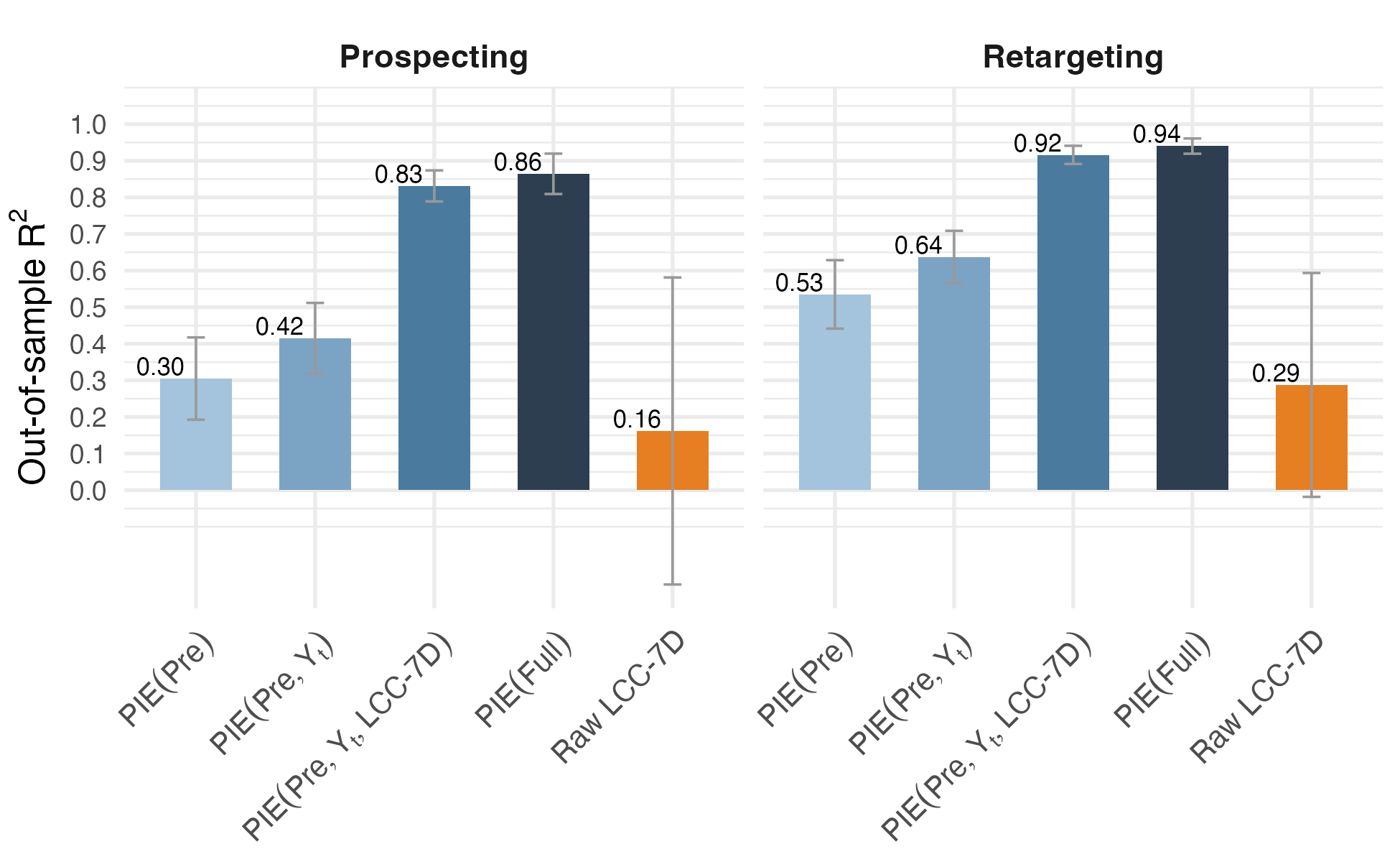} 
    \label{fig:pie_main_results_prospecting}
\begin{minipage}{0.98\textwidth} 
    \footnotesize
    \textit{Notes:} This figure compares the out-of-sample $R^2$ to predict ICPD ($\hat{\psi}_{r}$), decomposed across campaigns that either primarily target new customers (``Prospecting'') or existing customers (``Retargeting''). The bars represent progressive additions of features to the PIE model, from $\text{PIE}(\text{Pre})$ (using only pre-determined features) to $\text{PIE}(\text{Full})$ (using all features). $\text{Raw LCC-7D}$ serves as a benchmark that divides the raw last-click 7-day conversion count by campaign spend and treats the result as if it were ICPD, with no model applied. The $R^2$ value is shown above each bar, and error bars represent one standard deviation from the mean $R^2$, estimated via bootstrap.
    \end{minipage}
\end{figure}

Figure~\ref{fig:pie_main_results_prospecting} separates the results according to the type of audience targeting used in each campaign. Campaigns were classified as ``prospecting'' if at least 70\% of the audience constituted new potential customers for the advertiser. Otherwise, we classified them as ``retargeting'' campaigns. The results demonstrate a similar pattern of predictive performance across the campaign types as observed in the earlier figures.\footnote{This classification serves as a rough approximation to infer the campaign type, as it is possible for an advertiser to target a mix of users with any given campaign. The qualitative findings are not overly sensitive to the 70\% threshold.}

\begin{figure}[t]
    \centering
    \caption{PIE model performance by industry vertical}
    \includegraphics[width=0.98\textwidth]{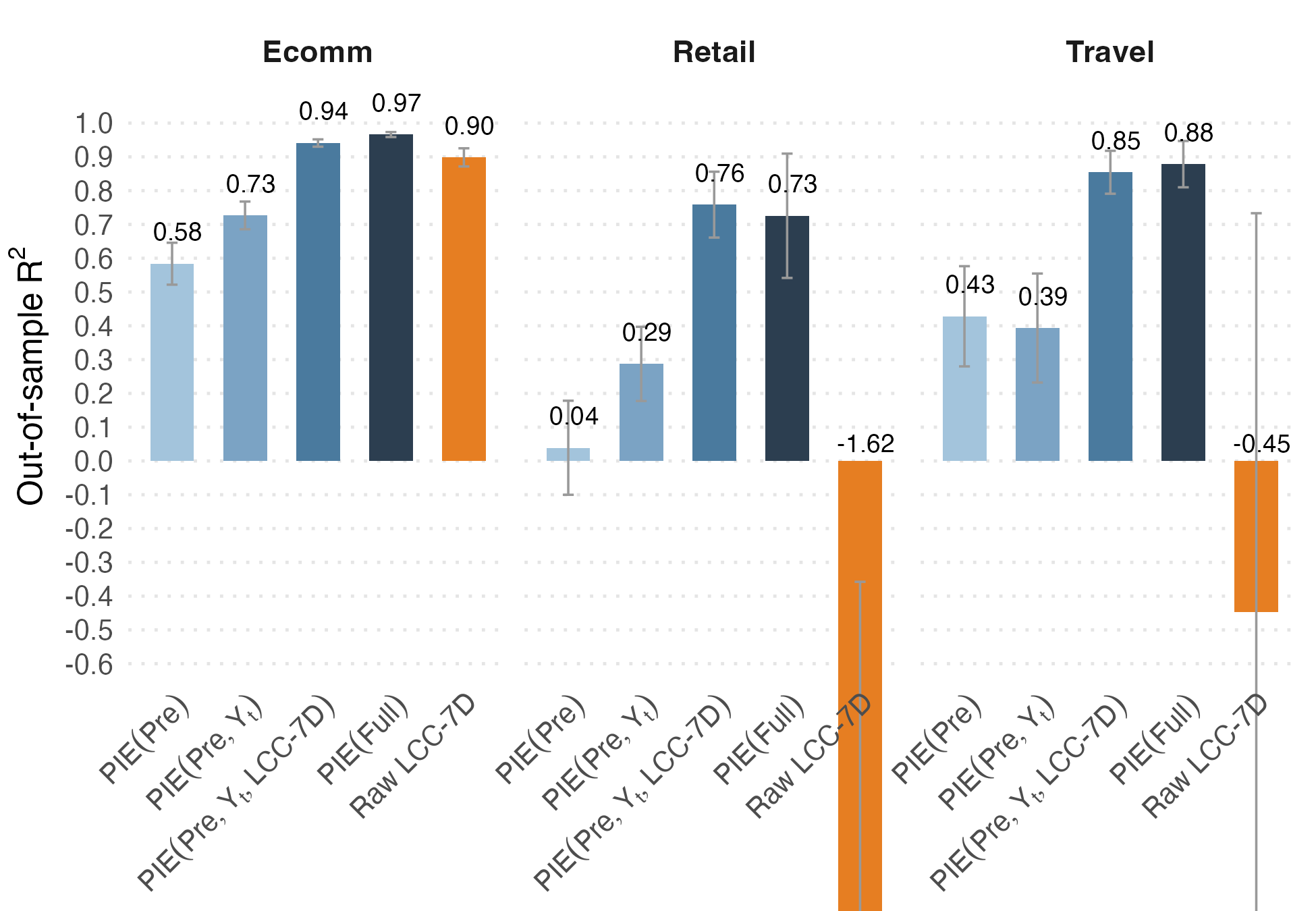} 
    \label{fig:pie_main_results_verticals}
\begin{minipage}{0.98\textwidth} 
    \footnotesize
    \textit{Notes:} This figure compares the out-of-sample $R^2$ to predict $\hat{\psi}_{r}$, decomposed across campaigns based on the product's industry vertical: e-commerce (``Ecomm''), retail, or travel. The bars represent progressive additions of features to the PIE model, from $\text{PIE}(\text{Pre})$ (using only pre-determined features) to $\text{PIE}(\text{Full})$ (using all features). $\text{Raw LCC-7D}$ serves as a benchmark that divides the raw last-click 7-day conversion count by campaign spend and treats the result as if it were ICPD, with no model applied. The $R^2$ value is shown above each bar, and error bars represent one standard deviation from the mean $R^2$, estimated via bootstrap. The $R^2$ values for $\text{Raw LCC-7D}$ in Retail and Travel are substantially negative, such that we have truncated the plot to maintain a readable scale for the positive $R^2$ values. The $R^2$ values shown by the text label (e.g., ``$-$1.62'') represent the true, untruncated values.    
    \end{minipage}
\end{figure}

Figure~\ref{fig:pie_main_results_verticals} presents the results organized by industry vertical of the advertiser for verticals with at least 200 RCTs. The figure shows that the PIE(Full) model outperforms the raw last-click metric, with the margin varying substantially across verticals. The ratio of mean LCC-7D per dollar to mean ICPD is 1.2 in e-commerce, 1.5 in retail, and 2.0 in travel, and OLS slopes of $\hat{\psi}_r$ on LCC-7D per dollar are 0.76, 0.63, and 0.48, respectively. In travel, approximately half of last-click conversions per dollar would have occurred without the ad. In e-commerce, where the bias is smallest, the raw metric still achieves reasonable out-of-sample performance ($R^2 = 0.90$), whereas in retail and travel, the systematic overstatement produces predictions worse than the sample mean.

\begin{figure}[t]
    \centering
    \caption{PIE model performance as a function of training sample size.\label{fig:robustness_all_sample_size_rsq}}
    \includegraphics[width=0.98\textwidth]{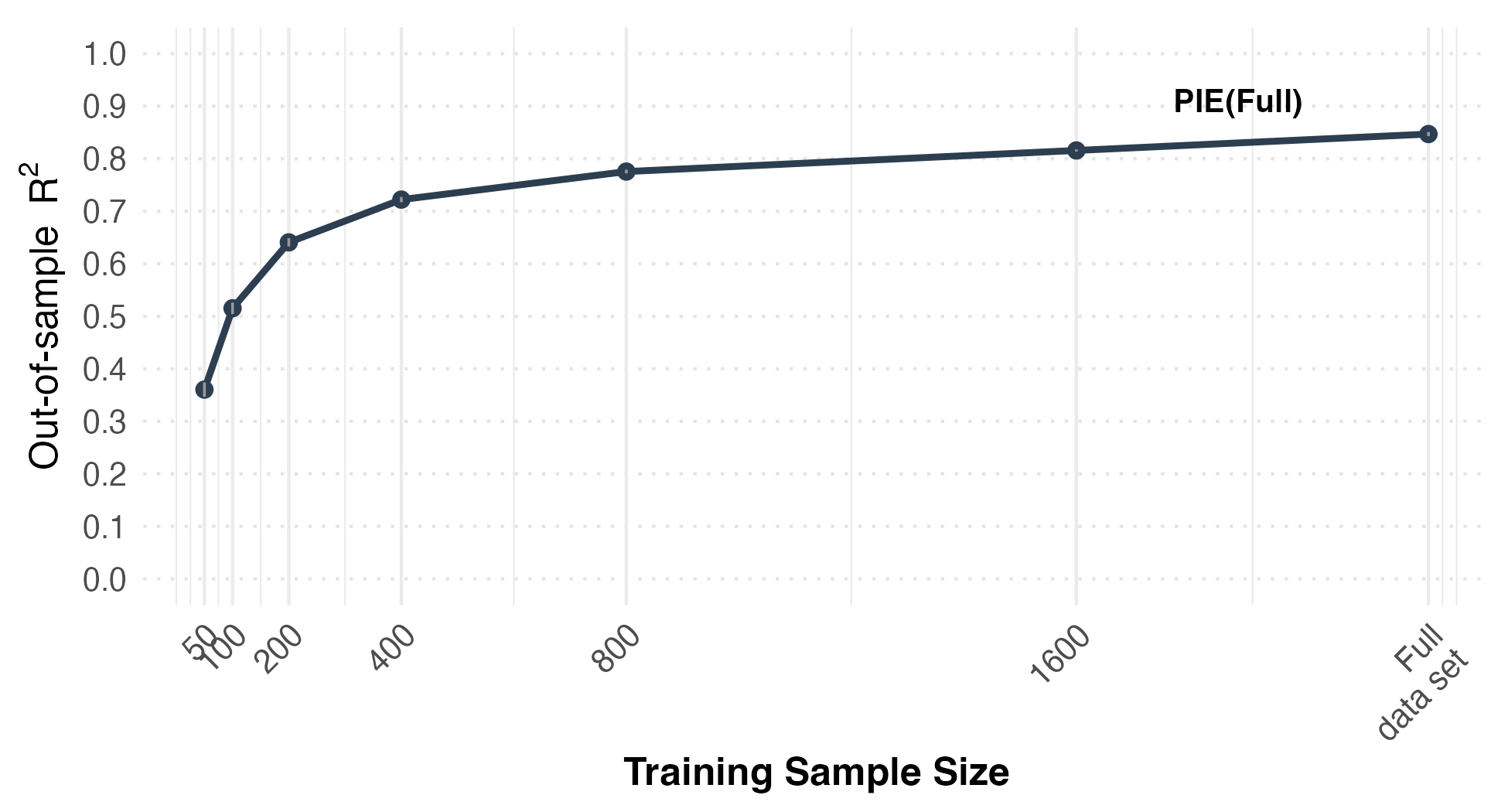} 
\begin{minipage}{0.98\textwidth} 
    \footnotesize
    \textit{Notes:} This figure illustrates the relationship between training sample size (number of RCTs) and the out-of-sample $R^2$ for the $\text{PIE(Full)}$ model. The $R^2$ values were obtained by repeatedly training the $\text{PIE(Full)}$ model on random subsamples of increasing size and evaluating its predictive performance.
    \end{minipage}
\end{figure}

Figure~\ref{fig:robustness_all_sample_size_rsq} investigates the trade-off between the quantity of RCTs and predictive performance. By repeatedly training the PIE(Full) model on random subsamples of increasing size and evaluating its $R^2$, we assess the model's performance with smaller datasets. We observe that, with a dataset of 800 RCTs (just over one-third the size of the full data), the $R^2$ is close to 80\%. However, the $R^2$ drops quickly as the sample size shrinks below 400 RCTs. 

In a real-world implementation, the platform must determine how to obtain a representative sample of RCTs while balancing the costs of running these experiments (e.g., the opportunity cost of withholding impressions from a control group). Furthermore, the platform must continually assess whether its current training sample has become stale, requiring new experiments to maintain model performance. We discuss these and related issues in Section~\ref{sec:conclusion}.

\subsection{Predictive Accuracy for New vs.~Existing Advertisers}\label{sec:results:newexisting}

One requirement for the practical application of the PIE model is its ability to provide accurate predictions for both existing and new advertisers. To test these two dimensions of performance, we implement two complementary cross-validation frameworks.

First, we ask: how well can the model predict for advertisers it has already seen? This addresses the model's utility in a stable environment with a known set of advertisers. To answer this, we employ a stratified $k$-fold cross-validation that trains the model on a subset of observations from every advertiser and tests it on a held-out subset from those same advertisers, thereby measuring its interpolative accuracy.\footnote{At the extreme, we could use a leave-one-RCT-out analysis to estimate the same quantity, but this is a relatively costly operation and in some testing we did not observe significant gains from doing so.}

Second, we ask: how well can the model generalize to new advertisers it has never seen before? This addresses the ``cold-start'' problem, a common challenge in marketing applications. To answer this more stringent question, we use a group $k$-fold cross-validation in which the holdout sets are composed of entirely unseen advertisers, meaning the model is trained on one cohort of advertisers and evaluated on its ability to predict for a completely separate cohort.

To implement these exercises within our data, we focus on the subset of advertisers with at least two RCTs, which yields a slightly smaller sample of 2,103 observations. In each exercise, we estimate the sampling variance using a non-parametric bootstrap across advertisers, sampling with replacement and performing each cross-validation procedure in its entirety for each bootstrap iteration.

\begin{figure}[t]
    \centering
    \caption{PIE Performance for Existing and New Advertisers}
    \includegraphics[width=0.98\textwidth]{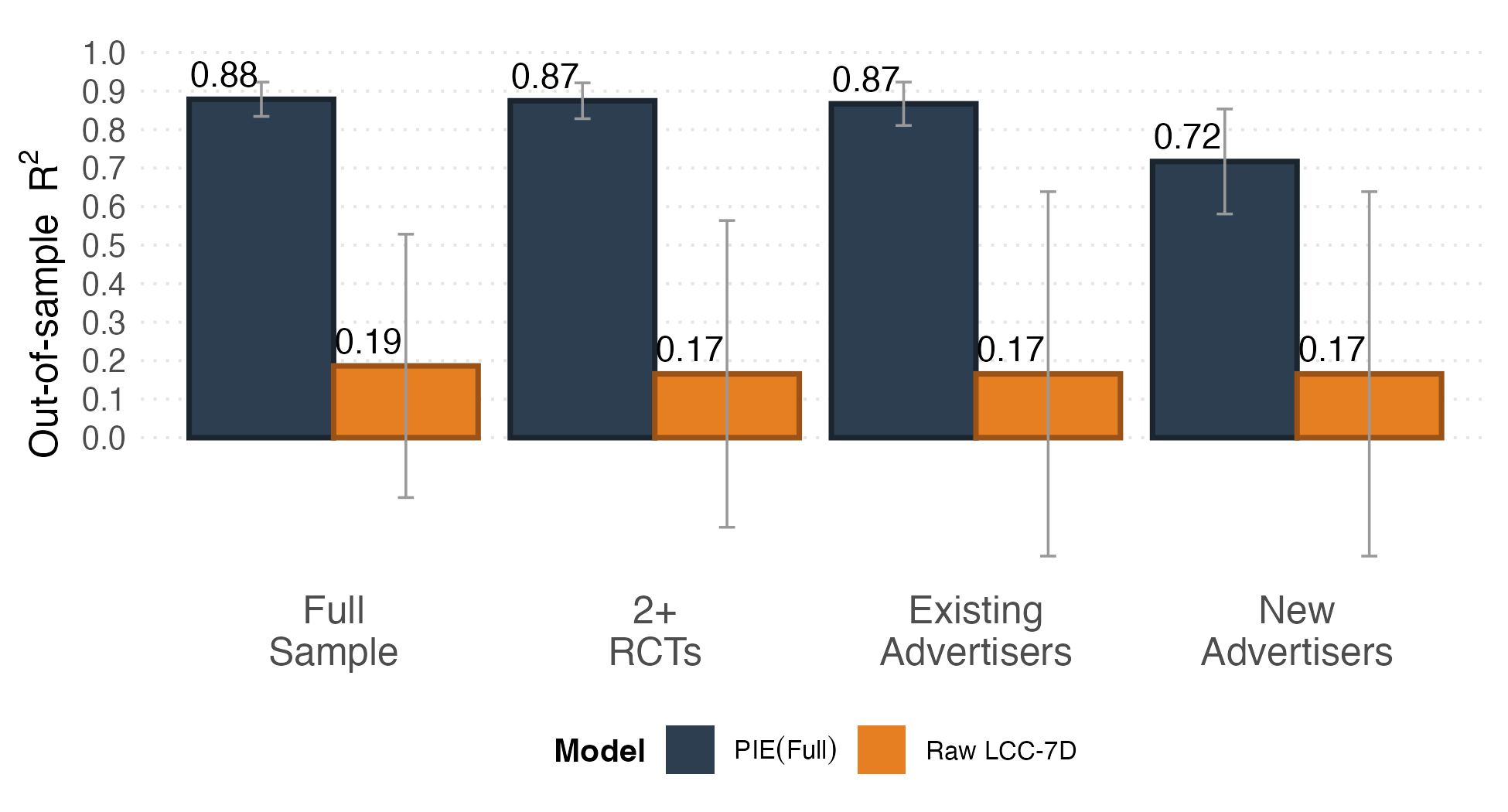} 
    \label{fig:robustness_all_rsq}
\begin{minipage}{0.98\textwidth} 
    \footnotesize
    \textit{Notes:} This figure presents the out-of-sample $R^2$ for the $\text{PIE(Full)}$ model and the $\text{Raw LCC-7D}$ benchmark across four conditions. The first two groups are based on simple $k$-fold cross-validation, using the $\text{Full Sample}$ (all 2,226 RCTs) and a restricted sample ($\text{2+ RCTs}$) used for the subsequent analyses, respectively. The remaining two groups of bars test the model's robustness using specialized cross-validation approaches: $\text{Existing Advertisers}$ measures interpolative accuracy using stratified $k$-fold CV (training and testing on different observations from the same advertisers). $\text{New Advertisers}$ measures extrapolative accuracy using group $k$-fold CV (training on one cohort of advertisers and testing on an entirely separate, unseen cohort). Error bars represent one standard deviation from the mean $R^2$.
    \end{minipage}
\end{figure}

These results are presented in Figure~\ref{fig:robustness_all_rsq}. On the left, we repeat the performance obtained by using PIE(Full) based on the full sample of 2,226 RCTs from Figure~\ref{fig:pie_main_results}. The second condition, ``2+ RCTs'', presents the predictive evaluation from 10-fold cross-validation using the narrower sample of advertisers with two or more RCTs. Thus, the performance of the PIE model on this slightly smaller sample of 2,103 RCTs is comparable to its performance on the full sample of 2,226 RCTs. Consistent with this finding, the raw last-click 7-day metric yields an $R^2$ that is 0.70 lower than the full PIE model.

PIE achieves the same performance as in the baseline when interpolating for ``Existing Advertisers.'' When extrapolating to ``New Advertisers'', the $R^2$ of the PIE model drops from 0.87 to 0.72 relative to the ``Existing Advertisers'' case. A reduction in predictive accuracy is expected when a model is used for extrapolation. It is difficult to assess whether this increase in error is ``large,'' in an absolute sense, since this depends on how different the new advertisers are relative to the training sample.

One useful point of comparison is the last-click 7-day metric. This metric yields the same $R^2 = 0.17$ regardless of whether the prediction is for new or existing advertisers, and performs substantially worse than PIE. This implies that an advertiser who had never run an RCT on the platform would benefit from evaluating their non-RCT campaigns using the full PIE model instead of the campaign's last-click 7-day conversions. 

\subsection{Predictive Accuracy for New Campaign Segments}\label{sec:results:segment}

The advertiser-level cross-validation in the preceding subsection tells us whether the model can generalize across \emph{firms}. In practice, however, a platform must also predict outcomes across \emph{segments} of ad campaigns that differ in creative strategy, campaign objectives, geography, or product vertical. To probe this second dimension of robustness, we conduct a hold-out-one-level experiment that asks a stricter question: How much accuracy is lost when the model must extrapolate from other segments to a focal segment it has never seen in training?

Although the advertiser-level cross-validation removes the target firm from the training set, the model may still learn from closely related advertisers. For example, if the training folds include Nike and Reebok, the model can transfer information related to their campaign features to predict effects for Adidas, even if Adidas is excluded from the training set. This overlap softens the cold-start problem. 

The segment-level experiment that follows is intentionally stricter. When we hold out one categorical level (e.g., advertisers belonging to the retail vertical), \emph{all} advertisers belonging to that level are excluded from training. The model must therefore extrapolate across both firms and segments, providing a sharper test of its ability to transfer knowledge to genuinely novel campaign contexts. By pairing two otherwise identical models---one trained inside the focal level and the other trained entirely outside it---we obtain an empirical distribution of ``extrapolation penalties'' that complements the advertiser-level metrics reported earlier.

We conduct this exercise using the following categorical features: advertiser size (small, medium, large), campaign year (using 2019 campaigns to predict 2020 campaigns), custom audience (broad or narrow), conversion optimization (offsite or onsite), prospecting or retargeting campaigns, and advertiser vertical (e-commerce, retail, and travel). 

Operationally, for each level~$\ell$ of a chosen campaign feature (e.g., prospecting campaigns), we first randomly divide all observations belonging to that level into two equal halves, each of size $R_{\ell}/2$. The first half is used to estimate a \emph{within-level} PIE model, while the second half is retained as a hold-out set unseen during training. A second model, the \emph{extrapolation} PIE model, is then trained on a sample of identical size $R_{\ell}/2$ drawn with replacement from the pool of observations outside level~$\ell$ (e.g., retargeting campaigns). Both models are evaluated on the \emph{same} hold-out set, such that any performance gap reflects only the cost of relying on observations from beyond the segment's focal level.\footnote{For example, consider the segmentation variable ``advertiser size'' $\in \{\text{small}, \text{medium}, \text{large}\}$. If the focal level is $\ell = \text{large}$, then we first split all campaigns with large advertisers into two groups, training on one and testing on the other, to produce our within-model evaluation. Next, we train the extrapolation model using an equal-sized training sample drawn randomly from the small and medium advertisers, and evaluate it using the same held-out test sample as with the within-model.} This entire split-train-test cycle is repeated 1,000 times with fresh random partitions, producing a distribution of $R^2$ for both scenarios at every level~$\ell$. 

\begin{figure}[t]
    \centering
    \caption{Extrapolation performance by advertiser size}
    \includegraphics[width=0.98\textwidth]{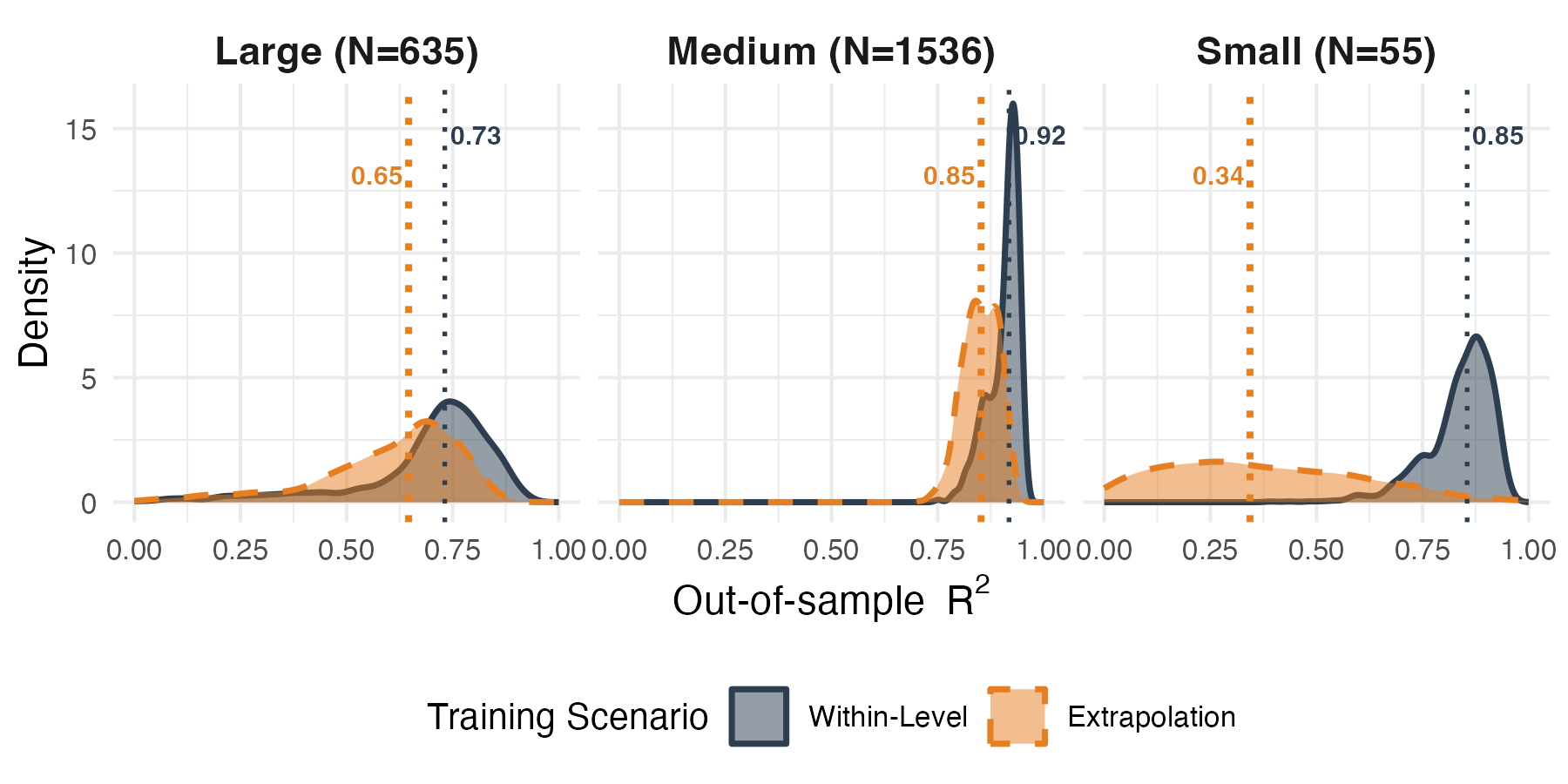} 
    \label{fig:rsq_density_Advertiser_size}
\begin{minipage}{0.98\textwidth}
    \footnotesize	
    \textit{Notes:} Empirical distributions of out-of-sample $R^2$ from the segment-level hold-out experiment, shown here for advertiser size. Each panel corresponds to one advertiser-size group (\emph{small}, \emph{medium}, \emph{large}), with separate density curves for models trained within the focal group (red) and for models trained by extrapolating from the remaining groups (blue). The distributions are based on 1,000 repeated random partitions of the data, providing a view of how predictive accuracy varies across resamples under the within- and extrapolation-training conditions. Vertical dotted lines indicate the median $R^2$ within each training scenario.
    \end{minipage}
\end{figure}

Figure~\ref{fig:rsq_density_Advertiser_size} visualizes these densities according to the focal segments defined by advertiser size, with figures for the other segmentation variables located in \ref{appendix:figs_tables}. We rely on a classification scheme that categorizes advertisers as ``small,'' ``medium,'' and ``large.'' Densities with lower within-scenario variance indicate that the model achieves fairly consistent within-segment predictive performance.

The extent to which the density for the extrapolation scenario (blue) is shifted to the left indicates the additional error induced by extrapolating from a model trained on data that completely lack the target level. In the left subfigure of Figure~\ref{fig:rsq_density_Advertiser_size}, we observe more consistency and little relative extrapolative error when the PIE model uses non-large advertisers to predict large advertisers. The penalty remains similarly low when we shift to predicting medium advertisers, which constitute the majority of observations in our sample. However, the extrapolative error increases sharply when we use a model trained on large and medium advertisers to predict ad effects for small advertisers, of which we have only 55 observations.

Table~\ref{tab:segment_extrapolation} on page~\pageref{tab:segment_extrapolation} reports the difference in the median out-of-sample $R^2$ across training scenarios for all the segmentation variables and their levels (see \ref{appendix:figs_tables} for additional plots). Across the segmentation variables, PIE predictions for extrapolations yield between 0.01 and 0.25 lower $R^2$ than their within-level PIE predictions.

\begin{table}[ht]
\centering
\caption{Out-of-sample $R^2$ performance: within-level versus extrapolation scenarios.\label{tab:segment_extrapolation}}
\centering
\fontsize{10}{12}\selectfont
    
\begin{tabular}{lccc}
\toprule
Segmentation Variable & Within-Level $R^2$ & Extrapolation $R^2$ & Difference\\
\midrule
\textbf{Conversion optimization} & \textbf{0.837} & \textbf{0.784} & \textbf{-0.053}\\
\hspace{1.5em}Opt Onsite Conversion & 0.854 & 0.824 & -0.031\\
\hspace{1.5em}Opt Offsite Conversion & 0.813 & 0.693 & -0.120\\
\textbf{Advertiser size} & \textbf{0.852} & \textbf{0.674} & \textbf{-0.178}\\
\hspace{1.5em}Large & 0.731 & 0.639 & -0.092\\
\hspace{1.5em}Medium & 0.918 & 0.852 & -0.066\\
\hspace{1.5em}Small & 0.855 & 0.303 & -0.552\\
\textbf{RCT Year} & \textbf{0.812} & \textbf{0.599} & \textbf{-0.213}\\
\hspace{1.5em}2020 & 0.812 & 0.599 & -0.213\\
\textbf{Prospecting vs. retargeting} & \textbf{0.824} & \textbf{0.767} & \textbf{-0.057}\\
\hspace{1.5em}Retargeting & 0.837 & 0.696 & -0.141\\
\hspace{1.5em}Prospecting & 0.806 & 0.796 & -0.009\\
\textbf{Vertical} & \textbf{0.825} & \textbf{0.691} & \textbf{-0.134}\\
\hspace{1.5em}Ecomm & 0.939 & 0.883 & -0.056\\
\hspace{1.5em}Retail & 0.724 & 0.622 & -0.102\\
\hspace{1.5em}Travel & 0.807 & 0.556 & -0.250\\
\textbf{Custom audience} & \textbf{0.706} & \textbf{0.559} & \textbf{-0.147}\\
\hspace{1.5em}Broad Audience & 0.736 & 0.612 & -0.124\\
\hspace{1.5em}Narrow Audience & 0.587 & 0.434 & -0.152\\
\bottomrule
\end{tabular}

    \begin{minipage}{0.94\textwidth}
        \footnotesize
        \textit{Notes:} Each row reports the median out-of-sample $R^2$ for the within-level and extrapolation training scenarios, along with their difference. Bolded values in the header rows represent weighted medians across levels within each segmentation variable.
    \end{minipage}
\end{table}

\section{Do Predictive Models Lead to the Same Decisions?} \label{sec:decision_making}

So far, we have focused on whether the ad effects estimates obtained from predictive models come close to those from RCTs. In practice, however, the estimates serve as a means to an end, which is to inform a managerial decision. As pointed out by \citet{FernandezLoriaProvost2022}, the task of estimating a causal effect is distinct from the task of making an optimal decision. An estimated effect with lower prediction error does not guarantee it will lead to better decisions, because the costs of different errors may be asymmetric and the magnitude of the error may matter differently depending on whether the estimate falls near a decision threshold. Therefore, to assess the practical value of PIE, we must evaluate it within a decision-making framework. What if a manager would make the same campaign decision regardless of whether they used a predictive model or an RCT? If this is the case, the practical implications of predictive models would be lessened despite their potential inaccuracy.

Our goal in this section is to assess the extent to which an advertiser using either last-click attribution or the PIE model would reach different conclusions about campaign effectiveness compared with using an RCT. We first formalize the advertiser's decision problem and describe how the use of different measurement methodologies can lead to different conclusions about a campaign's success. We then quantify these differences by estimating the probability that the RCT-based and model-based decisions would \emph{disagree}.

\subsection{Disagreement Probabilities}

How should we think about managerial decisions related to advertising campaigns? We adopt a perspective common in industry practice: marketers evaluate campaign performance based on incremental conversions per dollar (ICPD), as defined in Section~\ref{sec:application:evaluation}.

We assume that a manager considers a campaign successful if its ICPD exceeds a threshold, $\text{ICPD}^{*}$. Otherwise, the campaign is deemed unsuccessful, and the manager would prefer to reallocate those marketing funds elsewhere. The threshold can be interpreted as the inverse of the break-even profit-per-conversion that would justify the campaign's cost. The key question is whether the manager would reach the same judgment (success or failure) when using model-based estimates as when using RCT-based ones.\footnote{The ICPD is simply the inverse of a ``Cost per X'' metric. Many marketers evaluate campaign success using measures such as Cost per Conversion (CPC), Cost per Acquisition (CPA), Cost per Completed View (CPCV), or Cost per Engagement (CPE) \citep{xaxis_2018}. These metrics relate campaign costs to outcomes and are typically interpreted as if they reflect incremental performance, even though they are often based on non-causal measures. See \url{https://growhackscale.com/glossary/cost-per-conversion-metric}, accessed October 12, 2025.}

To illustrate, consider a manager who sets an advertising budget of \$100,000 and an ICPD threshold of $\text{ICPD}^{*}=0.01$, corresponding to one incremental conversion per \$100 spent. If an RCT shows that the campaign generated 500 incremental conversions, the observed $\text{ICPD}_{r}=500/100{,}000=0.005$, which falls short of the threshold. The manager would thus conclude the campaign failed.

Now suppose instead that the manager relied on a model that estimated 2,000 incremental conversions, implying $\text{ICPD}^{\text{pred}}_{r}=0.02$. Based on this model, the campaign would appear successful. The RCT and model therefore lead to opposing judgments (a disagreement).

This particular case represents a false positive, or Type I error: the model concludes that the campaign was successful when the RCT indicates otherwise. If the opposite occurred---the RCT judged the campaign successful but the model did not---the disagreement would represent a false negative, or Type II error. When both methods classify the campaign on the same side of the threshold, we say they agree.

We estimate the disagreement probability between the RCT and two models, the PIE(Full) and last-click 7-day. A high disagreement probability implies that model-based evaluations would often lead advertisers to make different campaign decisions than they would under the RCT benchmark, potentially misallocating spend despite small average estimation errors. Conversely, a low disagreement probability means that, even if model estimates are biased, advertisers would typically reach the same go/no-go decisions as they would with an RCT, making the model a practically reliable substitute for experimental measurement.

For a given value of $\text{ICPD}^{*}$, we calculate the disagreement probability over the set of campaigns as the sum of the Type~I and Type~II error probabilities:
\begin{align*}
\text{Disagreement}(\text{ICPD}^{*}) 
&= \frac{1}{R} 
   \sum_{r \in \mathcal{R}} 
   \underbrace{\mathbbm{1}\{\text{ICPD}_{r} \leq \text{ICPD}^{*}\} 
               \cdot 
               \mathbbm{1}\{\text{ICPD}_{r}^{\text{pred}} > \text{ICPD}^{*}\}}_{\text{Type I (false positive)}} \\
&\quad + 
   \underbrace{\mathbbm{1}\{\text{ICPD}_{r} > \text{ICPD}^{*}\} 
               \cdot 
               \mathbbm{1}\{\text{ICPD}_{r}^{\text{pred}} \leq \text{ICPD}^{*}\}}_{\text{Type II (false negative)}} .
\end{align*}
Because our interest lies in whether a manager would make the same campaign-level decision when relying on model-based estimates rather than the RCT benchmark, each campaign represents a single decision instance. We therefore compute the disagreement probability by giving every campaign equal weight.

To define the relevant range of $\text{ICPD}^*$ thresholds, we normalize the ICPD values for campaigns according to the median ICPD among campaigns in the same industry vertical and with outcomes at the same event funnel position. We do so because campaigns optimizing for upper-funnel events (e.g., page views) typically exhibit higher ICPD than those optimizing for lower-funnel conversions (e.g., purchases), and there is natural variation in ICPD across verticals. We examine thresholds in the range of 50\% to 150\% of each median, helping to ensure the thresholds we consider are economically relevant within each campaign segment.

We implement this approach using a non-parametric bootstrap, sampling observations with replacement and then training the PIE(Full) model on each sample (using 10-fold cross-validation). Then we estimate the disagreement probabilities point-wise at each $\text{ICPD}^{*}$, with the median normalization computed separately for each bootstrap replication. This process is repeated 1{,}000 times. 

\subsection{Results}

\begin{figure}[htbp]
    \centering
    \caption{Disagreement Probability}
    \includegraphics[width=0.98\textwidth]{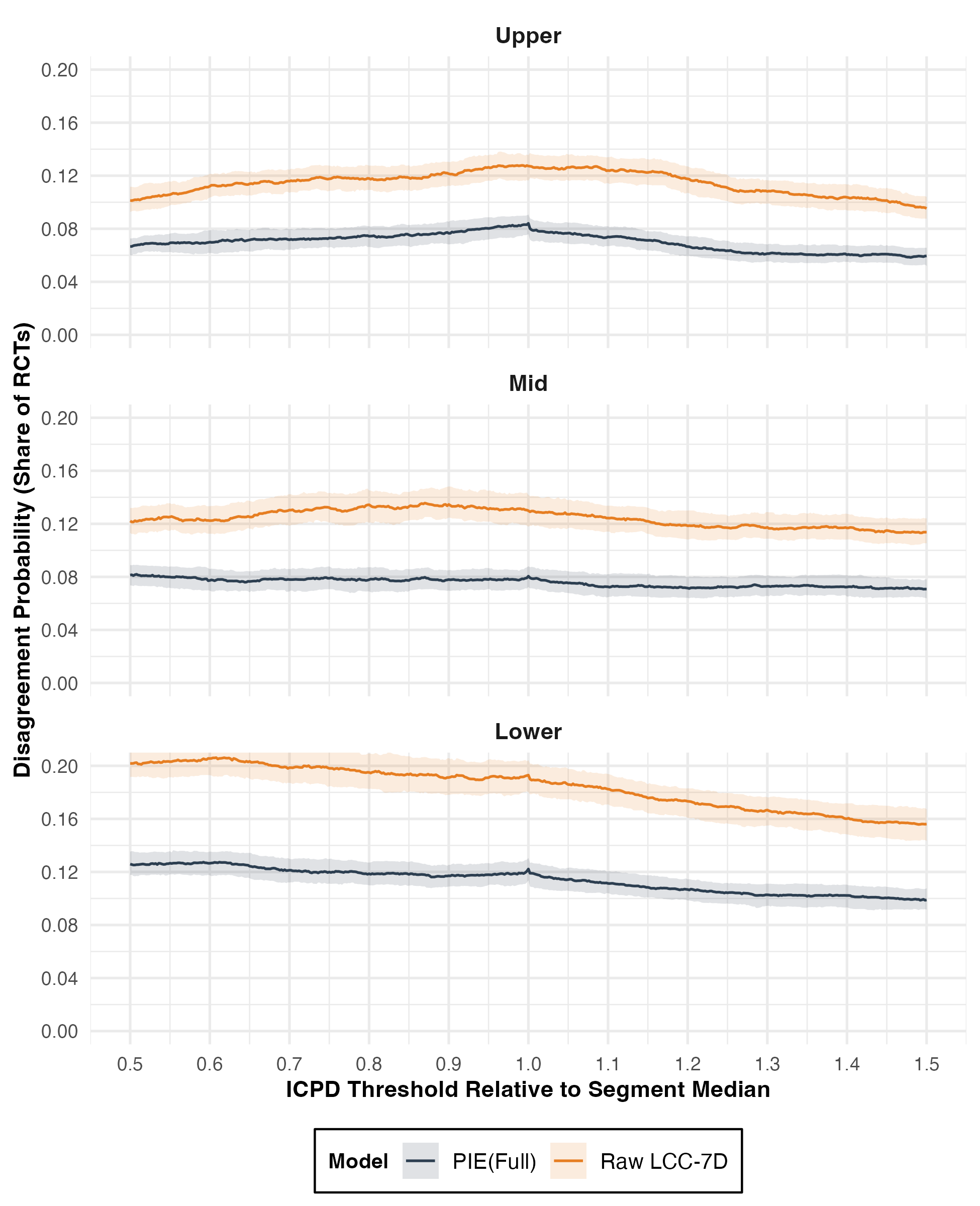} 
    \label{fig:disagreement_bands_by_funnel}
\begin{minipage}{0.98\textwidth}
    \footnotesize
    \textit{Notes:} Plots the share of RCTs for which an advertiser's decision about campaign success would differ when using model-based estimates of incremental conversions per dollar (ICPD) instead of the RCT benchmark, separately by event funnel. Decisions are based on whether ICPD exceeds a threshold defined relative to each segment's (event-funnel and vertical) median ICPD. The curves report the median disagreement probability across bootstrap repetitions for the PIE(Full) model (dark blue) and the raw last-click conversion 7-day metric (orange). Shaded regions denote the interquartile range across bootstrap repetitions.
    \end{minipage}
\end{figure}

Figure~\ref{fig:disagreement_bands_by_funnel} plots the median disagreement probabilities by funnel level when using either the PIE(Full) model vs.~RCT (dark blue) or the raw last-click 7-day vs.~RCT (orange), with the shaded regions corresponding to the inter-quartile range. The value of one on the x-axis corresponds to the median value within any given campaign's segment (vertical-funnel level). 

To interpret the figure, suppose that an advertiser sets a threshold of $\text{ICPD}^{*} = 1$. At such a threshold, the ICPD derived from a last-click 7-day attribution metric would ``disagree'' with the ICPD from an RCT in approximately 19.0\% of the lower-funnel campaigns, compared with approximately 12\% when using the PIE model. This finding for last-click attribution should not be generalized beyond the present setting, as the degree of attribution bias is likely platform-specific. In search advertising, clicks arise from latent purchase-intent prompting a query. In contrast, display ad clicks may reflect interest-based targeting and may therefore exhibit a different relationship between click behavior and true incremental effects.

\begin{figure}[htbp]
    \centering
    \caption{Disagreement Probability by Type I and II Errors}
    \includegraphics[width=0.98\textwidth]{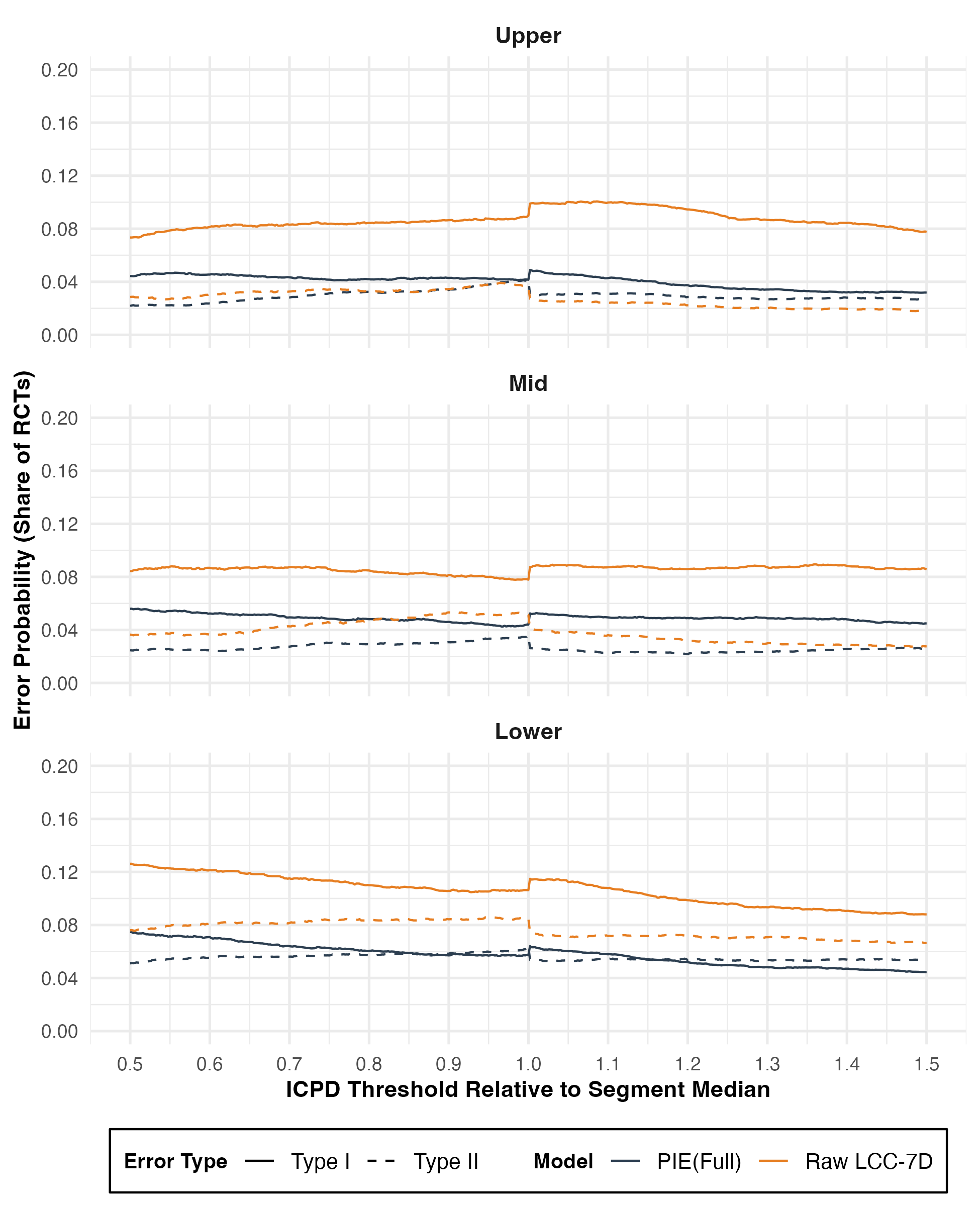} 
    \label{fig:disagreement_type_errors_by_funnel}
\begin{minipage}{0.98\textwidth}
    \footnotesize
    \textit{Notes:} Decomposes the disagreement probability from Figure~\ref{fig:disagreement_bands_by_funnel} into Type I errors (solid lines: false positives, where the model predicts ICPD above threshold but the RCT benchmark is below) and Type II errors (dashed lines: false negatives, where the model predicts ICPD below threshold but the RCT benchmark is above), separately by event funnel position. Thresholds are defined relative to each segment's median ICPD. The curves report the median error probability across bootstrap repetitions for the PIE(Full) model (dark blue) and the last-click conversion 7-day metric (orange).
    \end{minipage}
\end{figure}

Figure~\ref{fig:disagreement_type_errors_by_funnel} decomposes the disagreement curves in Figure~\ref{fig:disagreement_bands_by_funnel} into their Type I and II error components (omitting the bootstrapped confidence bands for clarity). The key observation is the asymmetry in error rates when relying on last-click 7-day conversions. The dashed orange line for last-click 7-day shows a lower Type II error rate but a substantially higher Type I error rate. This pattern arises because last-click 7-day metrics tend to be upwardly biased: by capturing conversions that occurred up to seven days after the impression and click-through, they systematically inflate ICPD and compress variation across campaigns. As a result, last-click rarely classifies a genuinely successful campaign as unsuccessful (low Type II error), but it frequently classifies unsuccessful campaigns as successful (high Type I error).

In contrast, the PIE model achieves low and balanced probabilities on both types of errors. If an advertiser is primarily focused on minimizing Type II error, then either the last-click 7-day metrics or PIE model perform similarly. However, if an advertiser cares similarly about both error types, then the PIE model outperforms last-click. 

Note that a manager may have different costs for committing either type of error. A false positive---incorrectly concluding that the campaign was a success---might lead the manager to implement future ad campaigns that are similarly unsuccessful, potentially wasting a significant amount of funds over time. A false negative might lead to the opposite outcome, causing an advertiser to hold back on further advertising spending even though it would have been the right decision. In this paper, we cannot assess the relative magnitudes of these costs. However, this information could be incorporated into the decision problem if the advertiser has access to such information.

\section{Conclusion} \label{sec:conclusion}

Advertisers and platforms need credible measures of incrementality, yet always-on RCTs are computationally costly to run at scale and observational proxies can be biased.

We propose Predicted Incrementality by Experimentation (PIE), a method to scale RCT causal effects to non-RCT campaigns. PIE uses a set of RCTs to learn a mapping from pre-determined (e.g., campaign configuration settings) and post-determined features (e.g., average outcomes, exposure and click behaviors, attribution counts) to the causal quantity of interest, and then applies that mapping to campaigns not run as RCTs. We view PIE first and foremost as a prediction, or supervised learning, problem. The emphasis is on accuracy in out-of-sample prediction, not identification of underlying mechanisms or user-level causal parameters.

Our framework clarifies why post-determined features---realized aggregates from the test group that most platforms already log---contain useful signal about incrementality. The same forces that shape causal effects, such as consumers' organic conversion rate and selection into exposure, also shape these post-determined features. PIE leverages this linkage by learning how multiple aggregate metrics can be combined to predict the causal estimand. 

Our primary contribution is to reframe non-RCT measurement as campaign-level prediction anchored by experiments and to formalize why post-determined features contain signal about incrementality. Because PIE relies only on test-group aggregates, it is feasible to deploy at scale with standard safeguards.

We present a proof-of-concept application of PIE using a sample of 2,226 ad experiments at Meta. PIE explains approximately 90\% of the out-of-sample variation in RCT effects and improves over common attribution benchmarks. PIE generalizes well within advertisers and, as expected, less well across segments that differ markedly from the training pool.

PIE is a general framework: causal effects can be mapped from RCTs to non-RCTs beyond the advertising domain, provided similar post-determined features that encode relevant structural parameters are available. 

Because PIE's accuracy depends on representativeness and overlap, an ad platform implementing PIE should curate a donor pool of RCTs that is both large and intentionally diverse across campaign characteristics (e.g., verticals, objectives, audience strategies, spend tiers, funnel events). In practice, this implies seeding RCTs where coverage is sparse (e.g., new objectives or formats), and excluding idiosyncratic tests that reflect one-off engineering configurations. PIE relies on the invariance of the relationship between features ($\mathbf X_r$) and causal targets ($\hat{\psi}_r$), so a platform must assess whether the distribution of features appears to shift over time and also ensure temporal coverage (rolling inclusion of recent campaigns). 

Finally, PIE is agnostic to precisely how the RCT outcomes are generated. We view platform-initiated experiments as particularly promising for generating PIE training data. In a ``shadow mode'' implementation, the platform continuously runs backend RCTs, invisible to advertisers, to fill gaps in the training pool. This helps ensure a steady supply of representative RCTs for model training. When the causal effects (labels) are noisier, they can be pooled with inverse-variance (meta-analytic) weights or stabilized via Bayesian hierarchical/shrinkage models and then fed into a supervised learning model. Choosing among these designs involves familiar trade-offs: the opportunity cost of withholding impressions, the engineering burden of managing a smaller number of shadow tests versus many smaller experiments, and the precision of each causal effect versus the representativeness of the donor pool. We view these as product and operations decisions rather than conceptual changes to PIE, and we leave a full treatment to future work.


\newpage
\lspace{1.1}
\bibliographystyle{chicago}
\bibliography{pie}

\clearpage

\appendix

\renewcommand{\thesection}{Appendix \arabic{section}} \makeatletter 
\renewcommand{\thesubsection}{\arabic{section}.\@arabic\c@subsection} \makeatother

\lspace{1.1}
\setcounter{equation}{0}
\setcounter{page}{1}
\setcounter{table}{0}
\setcounter{figure}{0}
\setcounter{section}{0}
\renewcommand{\thepage}{A-\arabic{page}}
\renewcommand{\thetable}{A-\arabic{table}}
\renewcommand{\thefigure}{A-\arabic{figure}}

\section{Notation} \label{app:notation}

This table summarizes the main notation used throughout the paper.

\begin{table}[h]
\centering
\caption{Key Notation}
\begin{tabular}{ll}
\toprule
\textbf{Symbol} & \textbf{Meaning} \\
\midrule
$r$ & Campaign/RCT index \\
$i$ & User index within campaign \\
$R$ & Total number of campaigns/RCTs \\
$N_r$ & Number of users in RCT $r$ \\
$N_{tr}$ & Number of users in RCT $r$ in the test group \\
$N_{cr}$ & Number of users in RCT $r$ in the control group \\
$Z_i^r$ & Random assignment indicator (1 = test, 0 = control) \\
$D_i^r$ & Exposure indicator (1 = exposed, 0 = not exposed) \\
$C_i^r$ & Click indicator (1 = clicked, 0 = did not click) \\
$U_i^r$ & User- and RCT-specific unobservable \\
$\alpha_r$ & Average baseline outcome (without exposure) in RCT $r$ \\
$\tau_r$ & Average treatment effect of exposure in RCT $r$ \\
$\gamma_r$ & Effect of user unobservable on baseline outcomes \\
$\delta_r$ & Causal effect of clicking on outcomes \\
$\phi_r$ & Exposure rate in test group of RCT $r$ \\
$\overline{Y}_{tr}$ & Average outcome in test group of RCT $r$ \\
$\overline{Y}_{cr}$ & Average outcome in control group of RCT $r$ \\
$\psi_r$ & Target causal estimand for PIE prediction  \\
$\mathbf{X}^{\mathrm{pre}}_r$ & Pre-determined features for RCT $r$ \\
$\mathbf{X}^{\mathrm{post}}_r$ & Post-determined features for RCT $r$ \\
$\text{ICPD}_r$ & Incremental conversions per dollar for RCT $r$ \\
$\text{LCC}_r(w)$ & Last-click conversions with attribution window $w$ \\
\bottomrule
\end{tabular}
\end{table}

\section{Data Details} \label{app:data_details}

This section provides additional details on the dataset of Meta ad experiments. Recall that we selected experiments with at least one million users assigned to test and retained outcome metrics with at least 5{,}000 conversions in the test group. These choices balance representativeness with statistical precision and computational feasibility. Initially, larger thresholds were used during preliminary analysis, after which we relaxed these thresholds to add experiments until the computational time required to extract the raw data became the binding constraint.

In November 2020, Meta (then Facebook) disclosed a bug affecting conversion metrics for about a year \citep{adexchanger_2020}. Our data were unaffected because we obtained them from an upstream source in the conversion measurement pipeline.

Figure~\ref{fig:dist_app} presents the empirical distribution of several experiment characteristics. The figure illustrates that we observe significant variation in the mix of experiments in our dataset. Table~\ref{tab:vertical_app} displays the distribution of RCTs by the industry vertical of the advertiser.

\begin{figure}[!ht]
\centering
\caption{Distribution of Experiment Characteristics}
\label{fig:dist_app}
\input{figures/distributions.tex}
\end{figure}

\begin{table}[!ht]
\centering
\caption{RCT Counts by Vertical}
\label{tab:vertical_app}

\begin{tabular}{lr}
\toprule
Vertical & N\\
\midrule
Ecomm & 616\\
Retail & 504\\
Travel & 350\\
EntMedia & 196\\
CPG & 121\\
Telecom & 105\\
Tech & 102\\
FinServ & 83\\
ProfServ & 65\\
Other & 36\\
Automotive & 34\\
Gaming & 14\\
\bottomrule
\end{tabular}

\end{table}

\section{Simulation Details} \label{Appendix:Simulation}

This appendix describes the simulation methodology used to evaluate PIE's predictive performance. Each simulation constructs potential outcomes for a set of campaigns $r \in \{1,\ldots,R\}$, each containing $N_r$ users. We use a linear probability model to generate user-level behaviors, ensuring that treatment effects are invariant to baseline conversion heterogeneity.

\subsection{Campaign-Level Parameter Generation}

For each campaign $r$, we draw the following parameters that govern user behavior and outcomes:

\paragraph{Exposure probability.} The probability of exposure $p_E^r$ is drawn from $p_E^r \sim U[p_E^{\text{low}}, p_E^{\text{high}}]$.

\paragraph{Baseline conversion probability and treatment effect.} We generate correlated baseline conversion probability $\alpha_r$ and treatment effect $\tau_r$ (measured in percentage points) using a Gaussian copula. This method allows us to independently control (1) the marginal distribution of each variable and (2) the correlation between them.

The procedure works as follows. First, we draw $(Z_1, Z_2)$ from a bivariate normal distribution with correlation $\rho_{\alpha\tau}$:
\begin{align}
\begin{pmatrix} Z_1 \\ Z_2 \end{pmatrix} \sim N\left(\begin{pmatrix} 0 \\ 0 \end{pmatrix}, \begin{pmatrix} 1 & \rho_{\alpha\tau} \\ \rho_{\alpha\tau} & 1 \end{pmatrix}\right)
\end{align}
This gives us two correlated normal random variables. Second, we apply the probability integral transformation to convert each normal variable to a uniform variable on $[0,1]$:
\begin{align}
U_1 = \Phi(Z_1), \quad U_2 = \Phi(Z_2)
\end{align}
where $\Phi(\cdot)$ is the standard normal cumulative distribution function. Because $\Phi(\cdot)$ is a monotonic transformation, $U_1$ and $U_2$ remain correlated with approximately the same rank correlation as $Z_1$ and $Z_2$. Moreover, by the probability integral transformation theorem, both $U_1$ and $U_2$ are uniformly distributed on $[0,1]$.

Finally, we transform these uniform variables to our desired marginal distributions using inverse transform sampling:
\begin{align}
\alpha_r &= \alpha^{\text{low}} + (\alpha^{\text{high}} - \alpha^{\text{low}}) \cdot U_1 \\
\tau_r &= \tau^{\text{low}} + (\tau^{\text{high}} - \tau^{\text{low}}) \cdot U_2
\end{align}
This maps each uniform variable to a uniform distribution on the specified range. These transformations preserve the dependence structure: $\alpha_r$ and $\tau_r$ inherit the correlation from $Z_1$ and $Z_2$, while having the desired marginal distributions (both uniform, but with different ranges). 

This approach gives us precise control over both the marginal distributions (the ranges from which $\alpha_r$ and $\tau_r$ are drawn) and their correlation, which would be difficult to achieve with simple bivariate sampling methods.

\paragraph{Effect of user propensity on outcomes.} The parameter $\gamma_r$ controls how strongly unobservable user propensity $U_i^r$ affects baseline conversion probability. We draw $\gamma_r \sim U[\gamma^{\text{low}}, \gamma^{\text{high}}]$.

\paragraph{Effect of clicks on conversions.} The parameter $\delta_r$ captures the causal effect of clicking on conversion probability (in percentage points). We draw $\delta_r \sim U[\delta^{\text{low}}, \delta^{\text{high}}]$.

\paragraph{Click probability.} The base click probability is drawn as $CP_r \sim U[CP^{\text{low}}, CP^{\text{high}}]$.

\subsection{Individual-Level Behavior Generation}

For each user $i$ in campaign $r$, we simulate the following behaviors:

\paragraph{User propensity.} We draw unobservable user propensity $U_i^r \sim \text{Uniform}(0,1)$.

\paragraph{Exposure.} Following equation (\ref{eq:exp_selection}), user $i$ is exposed if and only if $U_i^r \geq 1 - p_E^r$:
\begin{align*}
D_i^r = \mathbbm{1}\{U_i^r \geq 1 - p_E^r\}
\end{align*}

\paragraph{Click behavior.} The probability that user $i$ clicks (conditional on exposure) depends on user propensity:
\begin{align*}
\Pr(C_i^r = 1 | D_i^r = 1) = CP_r + \beta_{CP} \cdot U_i^r
\end{align*}
where $\beta_{CP}$ is a fixed slope parameter we specify in our simulations. Users click only if exposed: $C_i^r = \mathbbm{1}\{\text{click}\} \cdot D_i^r$.

\paragraph{Potential outcomes.} Using a linear probability model, we generate outcomes for the treatment and control conditions. The underlying probabilities are:
\begin{align*}
\text{Prob}_i^{r,T} &= \alpha_r + \gamma_r \cdot U_i^r + \tau_r \cdot D_i^r + \delta_r \cdot C_i^r \\
\text{Prob}_i^{r,C} &= \alpha_r + \gamma_r \cdot U_i^r
\end{align*}
We set our simulation parameters such that these probabilities are bounded within $[0,1]$.
Finally, we draw binary outcomes:
\begin{align*}
Y_{ti}^r &\sim \text{Bernoulli}(\text{Prob}_i^{r,T}) \\
Y_{ci}^r &\sim \text{Bernoulli}(\text{Prob}_i^{r,C})
\end{align*}

\paragraph{Last-click conversions.} We define a last-click conversion as $LCC_i^r = Y_{ti}^r \cdot C_i^r$, indicating that user $i$ both clicked and converted in the treatment condition.

\subsection{Aggregation and Sample Construction}

We aggregate individual-level data at the campaign level, yielding for each campaign $r$ the number of conversions in each group ($Y_{tr}, Y_{cr}$), clicks ($C_{tr}\geq 0$, $C_{cr}=0$), last-click conversions ($LCC_{tr}\geq 0$, $LCC_{cr}=0$), and the ad-exposure rate ($\phi_{tr}\geq 0$, $\phi_{cr}=0$).

\textbf{Sample splitting.} To avoid overfitting in small samples, we implement sample splitting within each campaign.\footnote{We thank Ryan Shyu for suggesting this.} For each campaign $r$, we randomly assign the $N_r$ users to two equally-sized samples. Sample 1 is used to calculate the observed incremental conversions $Y_{tr} - Y_{cr}$, which we use to calculate $\hat{\psi}_r$, the ATTs that are the dependent variable in the PIE regression. Sample 2 is used to calculate the predictors ($Y_{tr}$, $\phi_{tr}$, $C_{tr}$, $LCC_{tr}$). This ensures that predictors and outcomes are calculated from independent samples, eliminating mechanical correlation due to sampling error.

\textbf{Train-test split.} After generating data for all campaigns and calculating both outcomes and predictors, we split campaigns into equally-sized training and test sets. We estimate PIE specifications on the training sample using OLS regression:
\begin{align}
    \hat{\psi}_r = \theta_0 + \theta_1 \cdot Y_{tr} + \theta_2 \cdot \phi_{tr} + \theta_3 \cdot C_{tr} + \theta_4 \cdot LCC_{tr} + \nu_r
\end{align}
where all right-hand side variables are observable from the treatment group alone. Using estimated coefficients from the training sample, we generate predictions for the test sample:
\begin{align}
    \hat{\psi}_r^{\text{PIE}} = \hat{\theta}_0 + \hat{\theta}_1 \cdot Y_{tr} + \hat{\theta}_2 \cdot \phi_{tr} + \hat{\theta}_3 \cdot C_{tr} + \hat{\theta}_4 \cdot LCC_{tr}
\end{align}

\subsection{Evaluation Metrics}

We evaluate prediction accuracy using $R^2$, which measures the proportional reduction in mean squared error relative to a naive baseline that predicts the mean treatment effect for every campaign:
\begin{align}
R^2 = 1 - \frac{\sum_r (\hat{\psi}_r - \hat{\psi}_r^{\text{PIE}})^2}{\sum_r (\hat{\psi}_r - \bar{\psi})^2}
\end{align}
where $\bar{\psi}$ is the mean ATT across test campaigns. Higher $R^2$ indicates that the model substantially outperforms the naive baseline; $R^2 = 0$ indicates no improvement over predicting the mean. 

We report $R^2$ for models with increasingly rich feature sets: total conversions only ($R^2_{y_t}$), adding exposure fraction ($R^2_{y_t, \phi}$), adding clicks ($R^2_{y_t, \phi, C}$), and the full model including last-click conversions ($R^2_{y_t, \phi, C, LCC}$). Each simulation is repeated 100 times with different random seeds, and we pool all test observations across repetitions to calculate a single $R^2$ for each scenario.

\subsection{Detailed Simulation Results}

\paragraph{Number and Size of RCTs}

Table~\ref{tab:sim_sample_size} shows how predictive performance improves with the number of RCTs available for training. With 50 RCTs, the full model achieves $R^2 = 0.260$, improving to 0.442 with 1,000 RCTs and 0.448 with 10,000 RCTs. This occurs because larger training samples allow the model to better distinguish signal from noise. The gains diminish at larger sample sizes, with the increment from 1,000 to 10,000 RCTs adding little to $R^2$. Note that the number of RCTs at which gains diminish is highly dependent on other parameters of the simulation. For example, the larger the treatment effect, the fewer RCTs are needed. 

The second panel examines RCT size, holding the number of RCTs fixed at 1,000. Increasing the sample size per RCT from 1,000 to 10,000 users improves $R^2$ from 0.190 to 0.442. This occurs because larger RCTs produce more precise estimates of $\psi_r$ and the predictor variables in the PIE models. 

\begin{table}[htbp]
\centering
\caption{PIE Performance: Varying Number and Size of RCTs}
\label{tab:sim_sample_size}
\begin{tabular}{l*{4}{c}}
\hline\hline
& \multicolumn{4}{c}{$R^2$ for PIE model using} \\
\cmidrule(lr){2-5}
Scenario & $y_t$ & $y_t, \phi$ & $y_t, \phi, C$ & $y_t, \phi, C, LCC$ \\
\hline
\multicolumn{5}{l}{\textit{Panel A: Varying Number of RCTs (size = 10,000)}} \\
50 RCTs & 0.385 & 0.385 & 0.331 & 0.260 \\
100 RCTs & 0.412 & 0.412 & 0.396 & 0.368 \\
1,000 RCTs & 0.446 & 0.446 & 0.444 & 0.442 \\
10,000 RCTs & 0.448 & 0.448 & 0.448 & 0.448 \\
\addlinespace
\multicolumn{5}{l}{\textit{Panel B: Varying Size of RCTs (N = 1,000)}} \\
1,000 users & 0.213 & 0.213 & 0.201 & 0.190 \\
5,000 users & 0.400 & 0.400 & 0.396 & 0.392 \\
10,000 users & 0.446 & 0.446 & 0.444 & 0.442 \\
\hline\hline
\end{tabular}
\begin{tablenotes}
\small
\item \textit{Notes:} Out-of-sample $R^2$ for PIE models. Panel A varies the number of RCTs while holding RCT size constant at 10,000 users. Panel B varies RCT size while holding the number of RCTs constant at 1,000. All scenarios assume full compliance ($p_E = 1.00$), $\alpha_r \sim U[0.13, 0.17]$, and $\tau_r \sim U[0.02, 0.06]$.
\end{tablenotes}
\end{table}

\paragraph{Heterogeneity in Baseline Conversion Rates and Treatment Effects}

Table~\ref{tab:sim_heterogeneity} examines how heterogeneity in $\alpha_r$ and $\tau_r$ affects the performance of PIE. Panel A varies baseline conversion probability ($\alpha$) heterogeneity while holding $\tau_r \sim U[0.02, 0.06]$. The full model's $R^2$ declines from 0.666 (low heterogeneity) to 0.282 (high). This occurs because greater heterogeneity in $\alpha_r$ increases the variance of $Y_{tr}$ without proportionally increasing our ability to predict $\psi_r$, making the naive mean-based baseline harder to beat, consistent with what we found in Section~\ref{sec:pie_performance}. 

Panel B varies treatment effect ($\tau_r$) heterogeneity with $\alpha_r \sim U[0.13, 0.17]$. The full model's $R^2$ increases from 0.152 (low heterogeneity) to 0.640 (high). When treatment effects vary substantially, the error of the naive baseline (mean prediction) increases, while the PIE model's predictions exploit the fact that $\tau$ is on both sides of equations~\ref{eq:pie_att_expected} and~\ref{eq:pie_att_expected_mod}.

\begin{table}[htbp]
\centering
\caption{PIE Performance: Varying Heterogeneity in $\alpha_r$ and $\tau_r$}
\label{tab:sim_heterogeneity}
\begin{tabular}{l*{4}{c}}
\hline\hline
& \multicolumn{4}{c}{$R^2$ for PIE model using} \\
\cmidrule(lr){2-5}
Scenario & $y_t$ & $y_t, \phi$ & $y_t, \phi, C$ & $y_t, \phi, C, LCC$ \\
\hline
\multicolumn{5}{l}{\textit{Panel A: Varying Heterogeneity of Baseline Conversion Rate ($\alpha_r$)}} \\
Low: $U[0.14, 0.16]$ & 0.669 & 0.669 & 0.668 & 0.666 \\
Moderate: $U[0.13, 0.17]$ & 0.449 & 0.449 & 0.447 & 0.445 \\
High: $U[0.12, 0.18]$ & 0.286 & 0.286 & 0.284 & 0.282 \\
\addlinespace
\multicolumn{5}{l}{\textit{Panel B: Varying Heterogeneity of Treatment Effects ($\tau_r$)}} \\
Low: $U[0.03, 0.05]$ & 0.161 & 0.161 & 0.156 & 0.152 \\
Moderate: $U[0.02, 0.06]$ & 0.440 & 0.440 & 0.439 & 0.437 \\
High: $U[0.01, 0.07]$ & 0.643 & 0.643 & 0.642 & 0.640 \\
\hline\hline
\end{tabular}
\begin{tablenotes}
\small
\item \textit{Notes:} Out-of-sample $R^2$ for PIE models under varying degrees of parameter heterogeneity. Panel A varies baseline conversion probability range while holding $\tau_r \sim U[0.02, 0.06]$ constant. Panel B varies treatment effect heterogeneity while holding $\alpha_r \sim U[0.13, 0.17]$ constant. All scenarios use 1,000 RCTs with 10,000 users each and full compliance ($p_E = 1.00$).
\end{tablenotes}
\end{table}

\paragraph{Correlation Between $\alpha_r$ and $\tau_r$}

Table~\ref{tab:sim_correlation} varies the correlation from $-0.9$ to $+0.9$ while holding the marginal distributions at $\alpha_r \sim U[0.13, 0.17]$ and $\tau_r \sim U[0.02, 0.06]$. Positive correlation improves $R^2$ substantially, increasing from 0.440 (zero correlation) to 0.886 (correlation = 0.90). This occurs because positive correlation amplifies the systematic signal in $Y_{tr}$: when campaigns with high baseline conversion rates also have large treatment effects, $Y_{tr}$ provides a strong signal about both components.

Conversely, negative correlation severely degrades performance, with $R^2$ falling to 0.016 at correlation = $-0.90$. Strong negative correlation means that in $Y_{tr} = \alpha_r + \tau_r + \varepsilon_{tr}$, $\alpha_r$ and $\tau_r$ nearly cancel each other out, leaving only noise as an explanatory variable.

\begin{table}[htbp]
\centering
\caption{PIE Performance: Varying Correlation Between $\alpha_r$ and $\tau_r$}
\label{tab:sim_correlation}
\begin{tabular}{l*{4}{c}}
\hline\hline
& \multicolumn{4}{c}{$R^2$ for PIE model using} \\
\cmidrule(lr){2-5}
Correlation & $y_t$ & $y_t, \phi$ & $y_t, \phi, C$ & $y_t, \phi, C, LCC$ \\
\hline
\multicolumn{5}{l}{\textit{Panel A: Positive Correlation}} \\
$\rho = 0.00$ & 0.443 & 0.443 & 0.442 & 0.440 \\
$\rho = 0.45$ & 0.660 & 0.660 & 0.659 & 0.658 \\
$\rho = 0.90$ & 0.888 & 0.888 & 0.887 & 0.886 \\
\addlinespace
\multicolumn{5}{l}{\textit{Panel B: Negative Correlation}} \\
$\rho = 0.00$ & 0.447 & 0.447 & 0.445 & 0.443 \\
$\rho = -0.45$ & 0.234 & 0.234 & 0.230 & 0.228 \\
$\rho = -0.90$ & 0.021 & 0.021 & 0.019 & 0.016 \\
\hline\hline
\end{tabular}
\begin{tablenotes}
\small
\item \textit{Notes:} Out-of-sample $R^2$ for PIE models under varying correlation between baseline conversion rates and treatment effects. Marginal distributions held constant at $\alpha_r \sim U[0.13, 0.17]$ and $\tau_r \sim U[0.02, 0.06]$. Correlation induced using a Gaussian copula. All scenarios use 1,000 RCTs with 10,000 users each and full compliance ($p_E = 1.00$).
\end{tablenotes}
\end{table}

\paragraph{Effect of Clicks on Conversions ($\delta_r$)}

Table~\ref{tab:sim_click_effects} examines scenarios where clicking has a direct causal effect on conversion probability, parameterized by $\delta_r$. 

Under low click probability ($CP_r \sim U[0, 0.1]$, $\beta_{CP}=0.1$), introducing $\delta_r$ heterogeneity has minimal impact, with $R^2$ remaining around 0.45--0.46 across all scenarios. When few users click, these features contain little information regardless of click effect variation.

Under high click probability ($CP_r \sim U[0, 0.3]$, $\beta_{CP}=0.5$), $R^2$ increases from 0.513 to 0.657 as $\delta_r$ heterogeneity increases. The incremental contribution of last-click conversions grows substantially with $\delta$ heterogeneity, adding only 0.001 under low heterogeneity but 0.057 under high heterogeneity.

\begin{table}[htbp]
\centering
\caption{PIE Performance: Varying Click Effects on Conversions ($\delta_r$)}
\label{tab:sim_click_effects}
\begin{tabular}{l*{4}{c}}
\hline\hline
& \multicolumn{4}{c}{$R^2$ for PIE model using} \\
\cmidrule(lr){2-5}
Scenario & $y_t$ & $y_t, \phi$ & $y_t, \phi, C$ & $y_t, \phi, C, LCC$ \\
\hline
\multicolumn{5}{l}{\textit{Panel A: Low Click Probability, $CP_r \sim U[0.00, 0.10]$, $\beta_{CP} = 0.10$}} \\
Low Het: $\delta_r \sim U[0.04, 0.06]$ & 0.448 & 0.448 & 0.451 & 0.448 \\
Mod Het: $\delta_r \sim U[0.03, 0.07]$ & 0.453 & 0.453 & 0.455 & 0.453 \\
High Het: $\delta_r \sim U[0.01, 0.09]$ & 0.458 & 0.458 & 0.460 & 0.464 \\
\addlinespace
\multicolumn{5}{l}{\textit{Panel B: High Click Probability, $CP_r \sim U[0.00, 0.30]$, $\beta_{CP} = 0.50$}} \\
Low Het: $\delta_r \sim U[0.04, 0.06]$ & 0.483 & 0.483 & 0.514 & 0.513 \\
Mod Het: $\delta_r \sim U[0.03, 0.07]$ & 0.511 & 0.511 & 0.537 & 0.548 \\
High Het: $\delta_r \sim U[0.01, 0.09]$ & 0.588 & 0.588 & 0.600 & 0.657 \\
\hline\hline
\end{tabular}
\begin{tablenotes}
\small
\item \textit{Notes:} Out-of-sample $R^2$ for PIE models when clicks have a causal effect on conversions. Panel A examines low baseline click probability; Panel B examines high click probability. Within each panel, we vary $\delta_r$ heterogeneity. Click probability (conditional on exposure) depends on user propensity: $\Pr(C_i^r = 1 | D_i^r = 1) = CP_r + \beta_{CP} \cdot U_i^r$. All scenarios use 1,000 RCTs with 10,000 users each, $\alpha_r \sim U[0.13, 0.17]$, $\tau_r \sim U[0.02, 0.06]$, and full compliance ($p_E = 1.00$).
\end{tablenotes}
\end{table}

\paragraph{Selective Exposure and Unobserved Heterogeneity}

We now examine how PIE performance depends on unobserved heterogeneity in baseline outcomes and selection into exposure. Recall from equation (\ref{eq:exp_selection}) that exposure is determined by $D_i^r = \mathbbm{1} \{U_i^r \geq 1 - p_r(Z_i^r)\}$, where $U_i^r \sim \text{Uniform}(0,1)$ governs selection into exposure. The parameter $\gamma_r$ controls how strongly $U_i^r$ affects baseline conversion probability. When $\gamma_r = 0$, user propensity determines exposure but has no effect on outcomes. When $\gamma_r > 0$, the same unobservable that determines exposure also affects baseline conversion rates.

\begin{enumerate}

\item[i.] \textbf{Effect of $\gamma_r$ Heterogeneity.} 

Panel A in Table~\ref{tab:sim_exposure} shows results with partial selection into exposure ($p_E \sim U[0.4, 0.6]$). The full model's $R^2$ declines from 0.568 (low $\gamma_r$ heterogeneity) to 0.353 (high heterogeneity). This is analogous to increasing heterogeneity in the baseline probability $\alpha_r$: greater heterogeneity in $\gamma_r$ increases the variance of $Y_{tr}$ without proportionally increasing our ability to predict $\psi_r$ because $\gamma_r$ is only on the right-hand side of equation~\ref{eq:pie_att_expected}.

\item[ii.] \textbf{Effect of Exposure Selection Under Different $\gamma_r$ Scenarios.} 

Panel B in Table~\ref{tab:sim_exposure} varies exposure from high ($p_E \sim U[0.7, 0.9]$) to highly selective ($p_E \sim U[0.1, 0.3]$) for the moderate $\gamma_r \sim U[0.10, 0.20]$ case. The full model's $R^2$ increases from 0.319 for high exposure probability, to 0.430 for medium exposure probability, and 0.737 for low exposure probability.

This substantial improvement occurs because as selection increases, only users with increasingly high $U_i^r$ are exposed, and $\phi_r$ becomes increasingly informative about user composition, which PIE can exploit relative to a prediction of the mean. The table shows that PIE's use of the exposure fraction ($\phi_r$) as a post-determined feature is helpful when the selection mechanism correlates with outcomes. When $\gamma_r > 0$, the unobservable that determines exposure also affects baseline outcomes, and $\phi_r$ becomes increasingly valuable as a proxy for user composition.

\item[iii.] \textbf{Click Features as Proxies for User Composition.}

Panel C demonstrates an important property of click-based features: they improve predictions even when clicks have no direct causal effect on conversions ($\delta_r = 0$). Under moderate unobserved heterogeneity ($\gamma_r \sim U[0.10, 0.20]$) and partial exposure ($p_E \sim U[0.4, 0.6]$), adding click features increases $R^2$ from 0.087 (using only $y_t$) to 0.196 (adding $\phi$), and further to 0.252 (adding clicks and last-click conversions).

This occurs because under partial exposure, clicks reveal information about user composition. Users with high $U_i^r$ are both more likely to be exposed and, through the $\gamma_r$ parameter, more likely to convert. Since click probability also depends on $U_i^r$ (via $\Pr(C_i^r = 1 | D_i^r = 1) = CP_r + \beta_{CP} \cdot U_i^r$), click features serve as a proxy for the unobserved user characteristics that drive both exposure and outcomes. This demonstrates that click features can be valuable predictors of treatment effects even in the absence of any causal click-to-conversion pathway.

\begin{table}[htbp]
\centering
\caption{PIE Performance: Heterogeneity $\gamma_r$ and Exposure Selection}
\label{tab:sim_exposure}
\begin{tabular}{l*{4}{c}}
\hline\hline
& \multicolumn{4}{c}{$R^2$ for PIE model using} \\
\cmidrule(lr){2-5}
Exposure Range & $y_t$ & $y_t, \phi$ & $y_t, \phi, C$ & $y_t, \phi, C, LCC$ \\
\hline
\multicolumn{5}{l}{\textit{Panel A: Varying Heterogeneity in the effect of $U_i$}} \\
Low Heterogeneity: $\gamma_r \sim U[0.14, 0.16]$ & 0.244 & 0.357 & 0.405 & 0.568 \\
Moderate Heterogeneity: $\gamma_r \sim U[0.10, 0.20]$ & 0.126 & 0.276 & 0.337 & 0.460 \\
High Heterogeneity: $\gamma_r \sim U[0.00, 0.30]$ & 0.022 & 0.210 & 0.285 & 0.353 \\
\addlinespace
\multicolumn{5}{l}{\textit{Panel B: Varying Exposure Selection}} \\
$p_E \sim U[0.70, 0.90]$ & 0.226 & 0.272 & 0.276 & 0.319 \\
$p_E \sim U[0.40, 0.60]$ & 0.120 & 0.284 & 0.289 & 0.430 \\
$p_E \sim U[0.10, 0.30]$ & 0.051 & 0.655 & 0.656 & 0.737 \\
\addlinespace
\multicolumn{5}{l}{\textit{Panel C: Click Features Without Click Treatment Effect ($\delta_r = 0$)}} \\
$\gamma_r \sim U[0.10, 0.20]$, $p_E \sim U[0.40, 0.60]$ & 0.087 & 0.196 & 0.193 & 0.252 \\
\hline\hline
\end{tabular}
\begin{tablenotes}
\small
\item \textit{Notes:} Out-of-sample $R^2$ under varying degrees of exposure selection and unobserved heterogeneity. Panels A and B use $\delta_r \sim U[0.03, 0.07]$. Panel C demonstrates that click features ($C$ and $LCC$) improve predictions even when clicks have no causal effect on conversions ($\delta_r = 0$), because they proxy for user composition under partial exposure. All scenarios use 1,000 RCTs with 10,000 users each, $\alpha_r \sim U[0.13, 0.17]$, and $\tau_r \sim U[0.02, 0.06]$.
\end{tablenotes}
\end{table}

\end{enumerate}

\section{Additional Extrapolation by Segment Figures} \label{appendix:figs_tables}

\begin{figure}[ht]
    \centering
    \caption{Extrapolation performance by advertiser vertical}
    \includegraphics[scale=0.95]{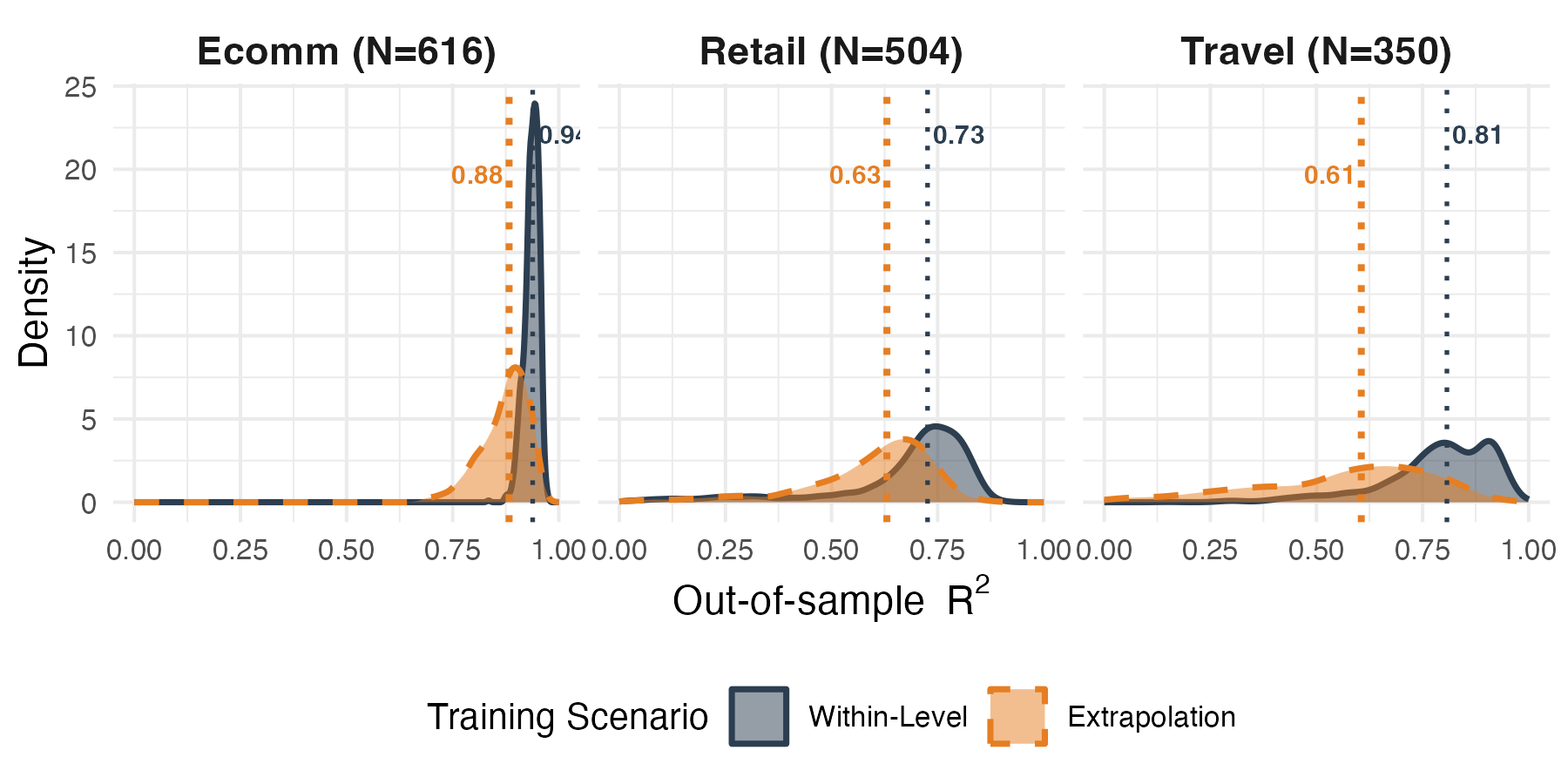} 
    \label{fig:rsq_density_Vertical}
\begin{minipage}{0.95\textwidth}
    \footnotesize
    \textit{Notes:} Empirical distributions of out-of-sample $R^2$ from the segment-level hold-out experiment, shown here for a subset of advertiser verticals. Each panel corresponds to a vertical (e-commerce, retail, travel), with separate density curves for models trained within the focal group (red) and for models trained by extrapolating from the remaining groups (blue). The distributions are based on 1{,}000 repeated random partitions of the data, providing a view of how predictive accuracy varies across resamples under the within- and extrapolation-training conditions. Vertical dotted lines indicate the median $R^2$ within each training scenario.
    \end{minipage}
\end{figure}

\begin{figure}[ht]
    \centering
    \caption{Extrapolation performance for prospecting vs.~retargeting campaigns}
    \includegraphics[scale=0.95]{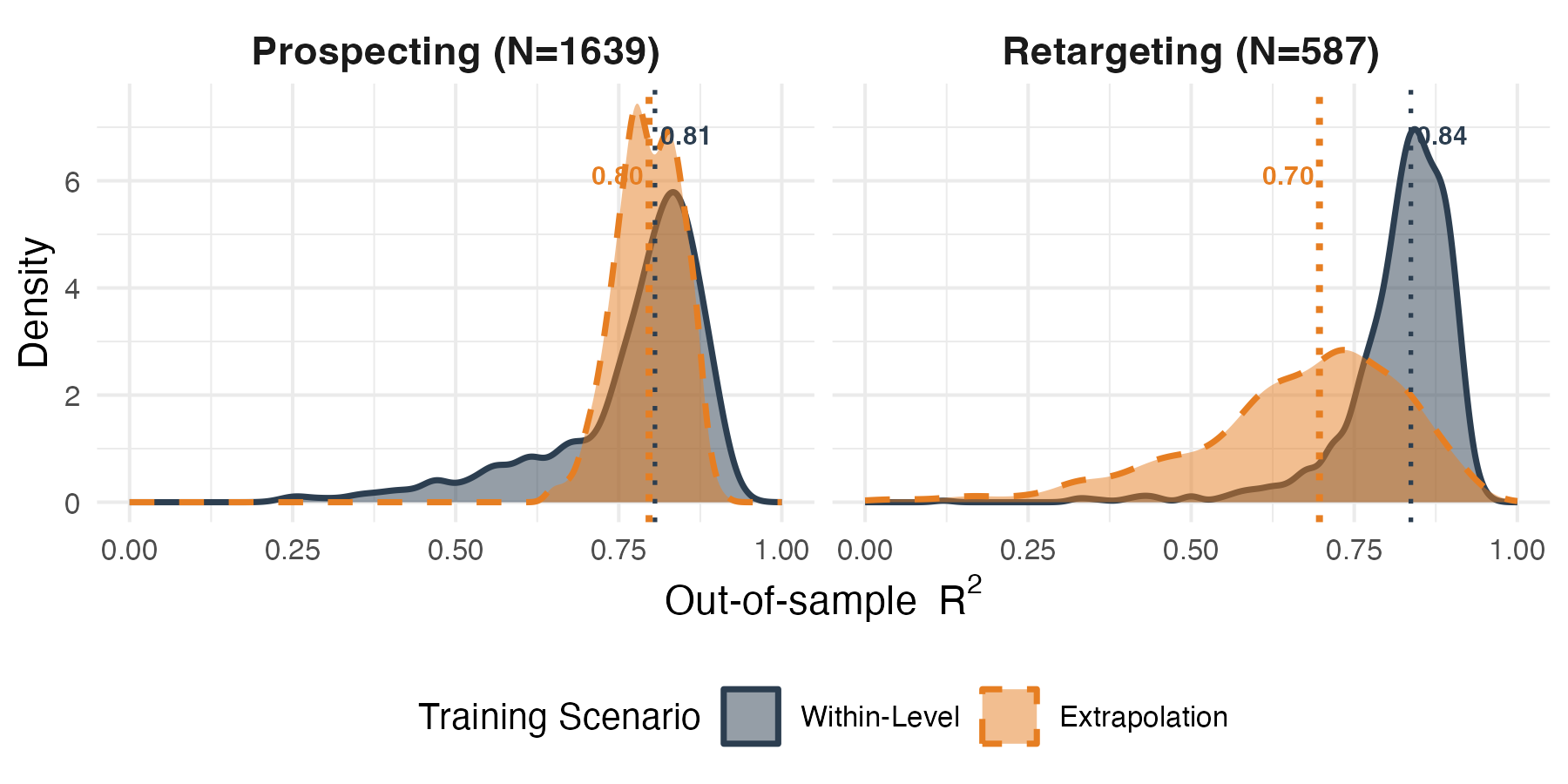} 
    \label{fig:rsq_density_prospecting_vs_retargeting}
\begin{minipage}{0.95\textwidth}
    \footnotesize
    \textit{Notes:} Empirical distributions of out-of-sample $R^2$ from the segment-level hold-out experiment, shown here based on whether the campaign is focused on reaching new potential customers (\textbf{Prospecting}) or aims to reach existing customers (\textbf{Retargeting}). In each panel, there are separate density curves for models trained within the focal group (red) and for models trained by extrapolating from the remaining groups (blue). The distributions are based on 1{,}000 repeated random partitions of the data, providing a view of how predictive accuracy varies across resamples under the within- and extrapolation-training conditions. Vertical dotted lines indicate the median $R^2$ within each training scenario.
    \end{minipage}
\end{figure}

\begin{figure}[ht]
    \centering
    \caption{Extrapolation performance for custom audience campaigns}
    \includegraphics[scale=0.95]{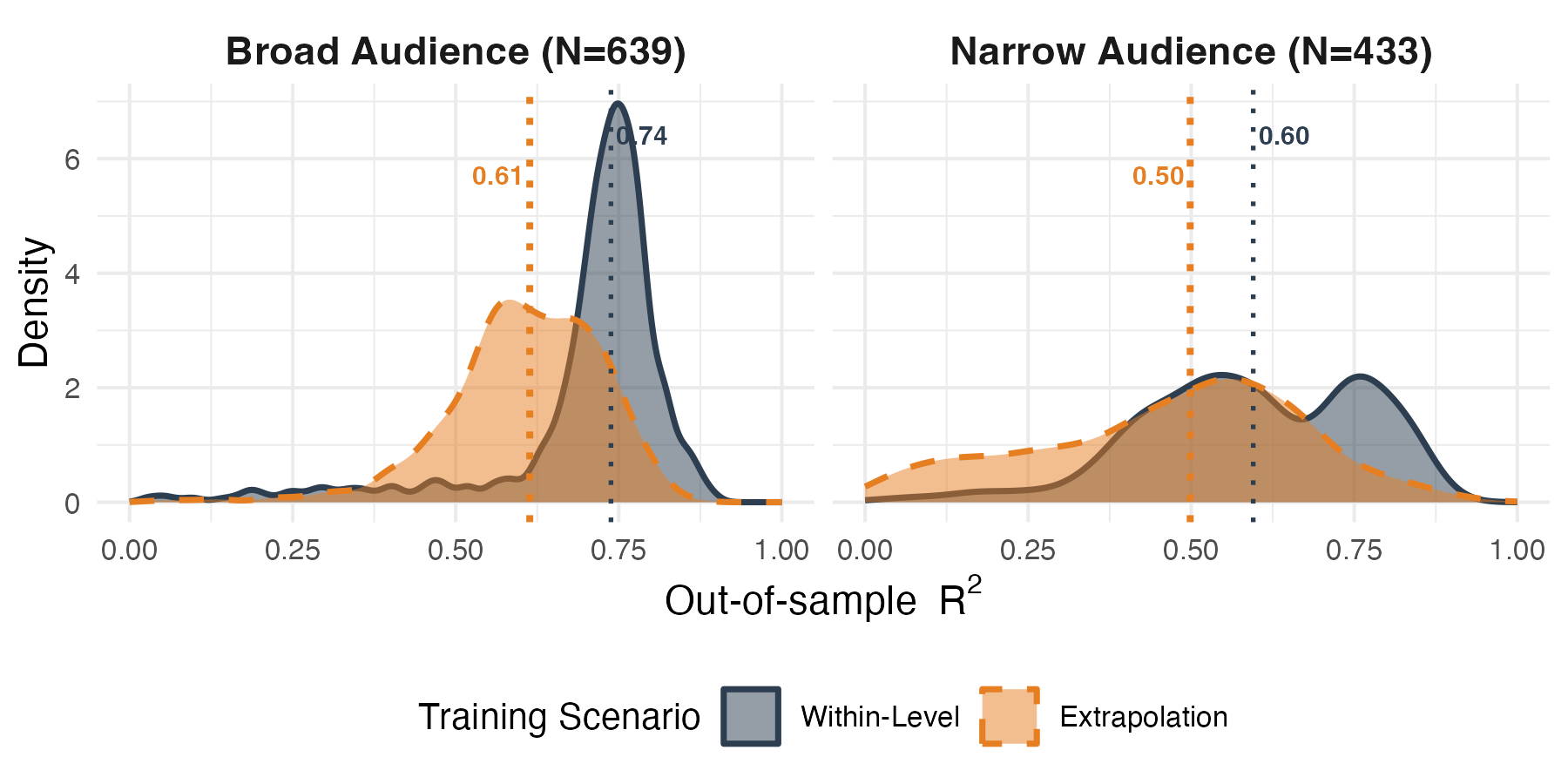} 
    \label{fig:rsq_density_Custom_audience}
\begin{minipage}{0.95\textwidth}
    \footnotesize
    \textit{Notes:} Empirical distributions of out-of-sample $R^2$ from the segment-level hold-out experiment, shown here based on whether the campaign targets a \textbf{narrow} custom audience or a \textbf{broad} custom audience. In each panel, there are separate density curves for models trained within the focal group (red) and for models trained by extrapolating from the remaining groups (blue). The distributions are based on 1{,}000 repeated random partitions of the data, providing a view of how predictive accuracy varies across resamples under the within- and extrapolation-training conditions. Vertical dotted lines indicate the median $R^2$ within each training scenario.
    \end{minipage}
\end{figure}

\begin{figure}[ht]
    \centering
    \caption{Extrapolation performance training on campaigns in 2019 to extrapolate to campaigns run in 2020}
    \includegraphics[scale=0.95]{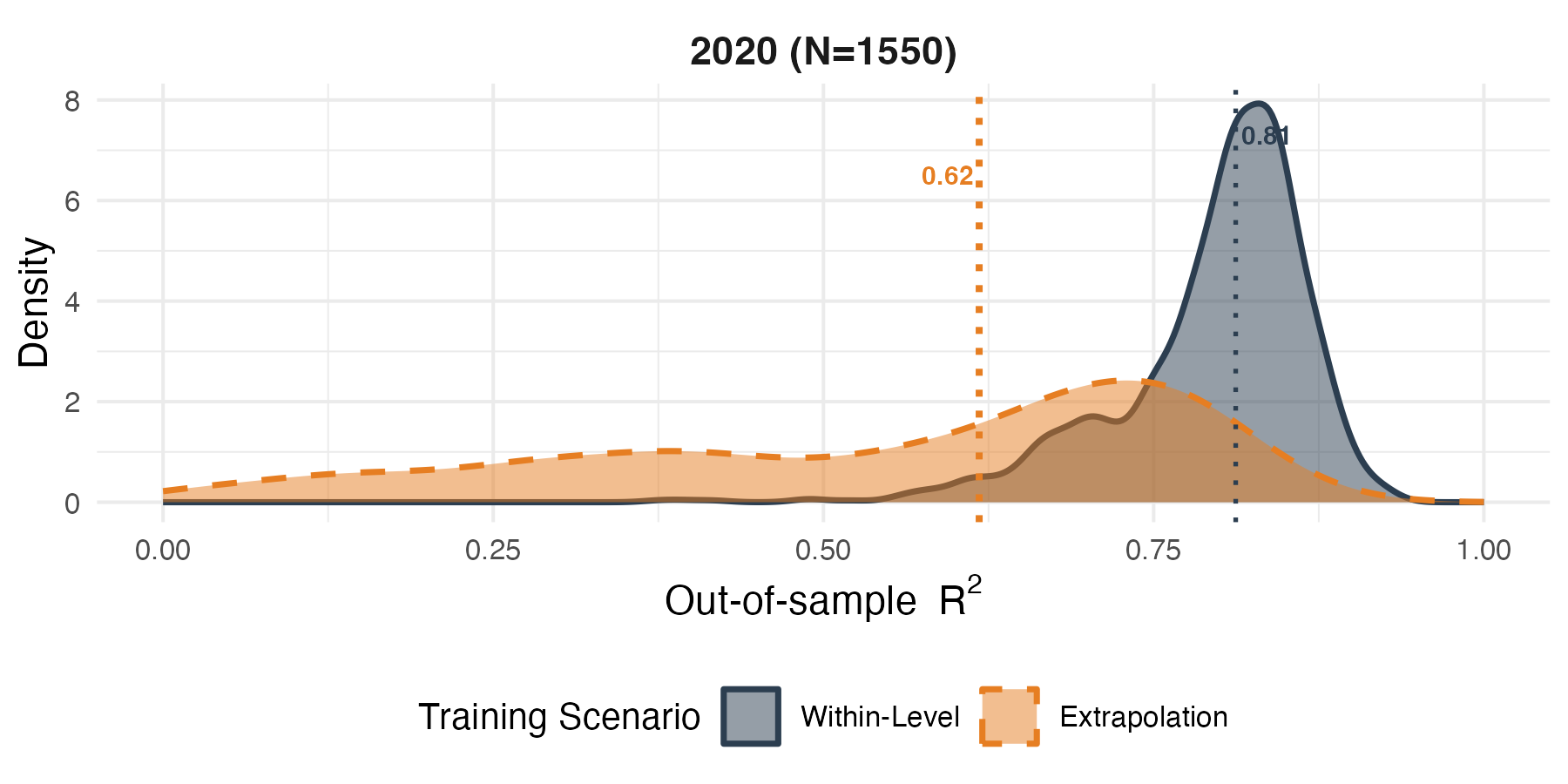} 
    \label{fig:rsq_density_RCT_Year}
\begin{minipage}{0.95\textwidth}
    \footnotesize
    \textit{Notes:} Empirical distributions of out-of-sample $R^2$ from the segment-level hold-out experiment. In this case, the within (red) trains a model on campaigns from 2020 and predicts on campaigns from 2020, whereas the extrapolation (blue) trains the model on campaigns from 2019 and predicts on campaigns from 2020. The distributions are based on 1{,}000 repeated random partitions of the data, providing a view of how predictive accuracy varies across resamples under the within- and extrapolation-training conditions. Vertical dotted lines indicate the median $R^2$ within each training scenario.
    \end{minipage}
\end{figure}

\begin{figure}[ht]
    \centering
    \caption{Extrapolation performance by type of conversion optimization (onsite vs.~offsite)}
    \includegraphics[scale=0.95]{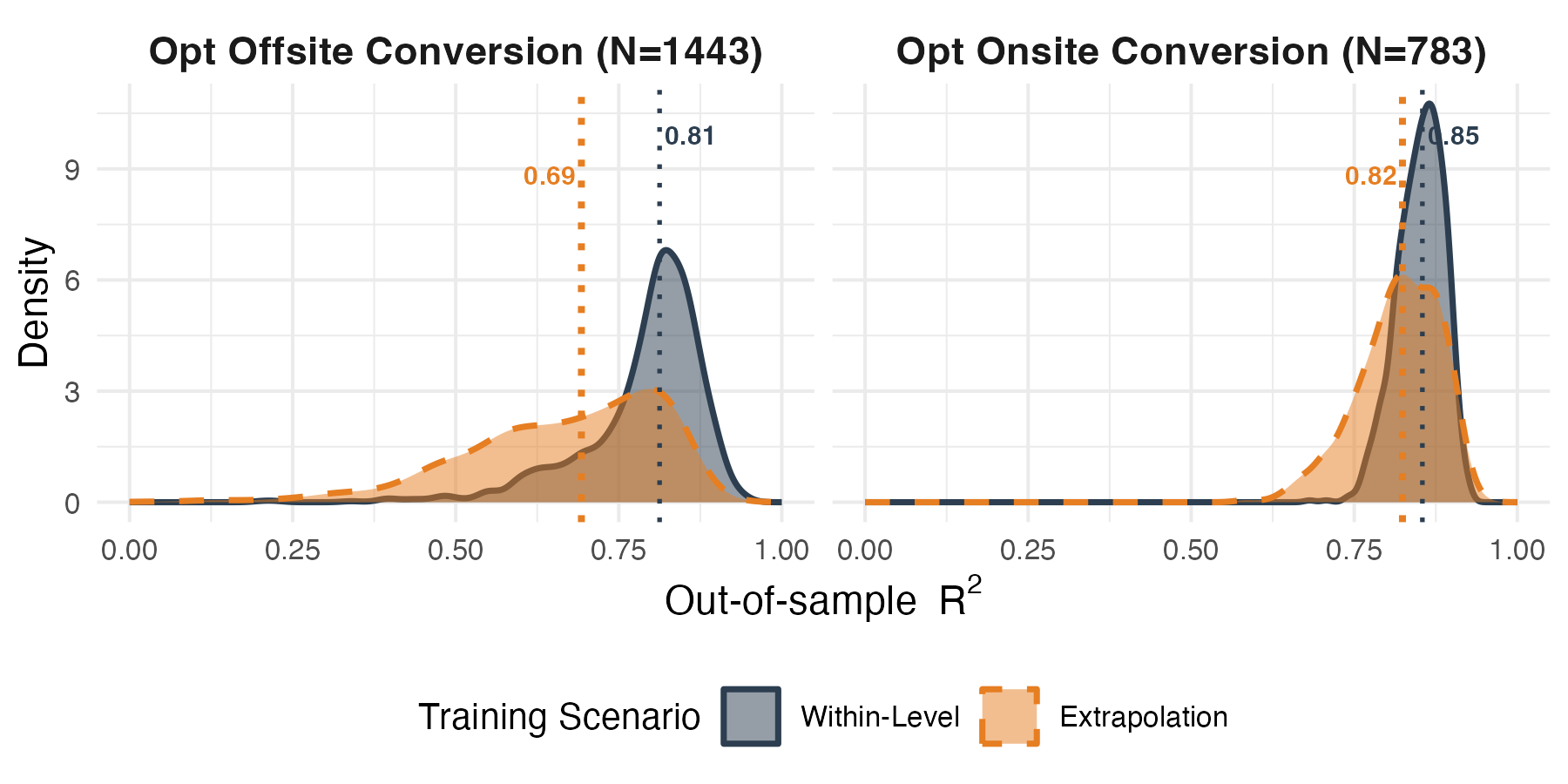} 
    \label{fig:rsq_density_Conversion_optimization}
\begin{minipage}{0.95\textwidth}
    \footnotesize
    \textit{Notes:} Empirical distributions of out-of-sample $R^2$ from the segment-level hold-out experiment, shown here based on whether the campaign is optimizing for offsite conversions or onsite conversions. In each panel, there are separate density curves for models trained within the focal group (red) and for models trained by extrapolating from the remaining groups (blue). The distributions are based on 1{,}000 repeated random partitions of the data, providing a view of how predictive accuracy varies across resamples under the within- and extrapolation-training conditions. Vertical dotted lines indicate the median $R^2$ within each training scenario.
    \end{minipage}
\end{figure}

\end{document}